\documentclass[twocolumn]{aastex63}
\usepackage[utf8]{inputenc}

\usepackage{natbib}
\usepackage{graphicx}
\usepackage{amsmath}
\usepackage{booktabs}
\usepackage{xspace}
\usepackage{hyperref}
\usepackage[T1]{fontenc}
\usepackage{lipsum}
\usepackage{comment}
\usepackage{xcolor}
\usepackage{multirow}
\usepackage{enumitem}
\usepackage{float}

\usepackage{IEEEtrantools}

\newcommand{\sbt}{\,\begin{picture}(-1,1)(-1,-3)\circle*{3}\end{picture}\ }
\usepackage{makecell}
\setcellgapes[t]{3pt}
\setcellgapes[b]{1pt}

\newcommand{\msun}{\ensuremath{M_{\odot}}}

\providecommand{\mj}{\ensuremath{\,M_{\rm J}}}

\begin{document}

\title{\texttt{ethraid}: A simple method for characterizing long-period companions using Doppler, astrometric, and imaging constraints}

\author[0000-0002-4290-6826]{Judah Van Zandt}
\affiliation{Department of Physics \& Astronomy, University of California Los Angeles, Los Angeles, CA 90095, USA}

\author[0000-0003-0967-2893]{Erik A Petigura}
\affiliation{Department of Physics \& Astronomy, University of California Los Angeles, Los Angeles, CA 90095, USA}

\begin{abstract}

We present \texttt{ethraid},\footnote{\cite{ethraidDOI}} an open source Python package designed to measure the mass ($m_c$) and separation ($a$) of a bound companion from measurements covering a fraction of the orbital period. \texttt{ethraid} constrains $m_c$ and $a$ by jointly modeling radial velocity (RV), astrometric, and/or direct imaging data in a Bayesian framework. Partial orbit data sets, especially those with highly limited phase coverage, are well-represented by a few method-specific summary statistics. By modeling these statistics rather than the original data, \texttt{ethraid} optimizes computational efficiency with minimal reduction in accuracy. \texttt{ethraid} uses importance sampling to efficiently explore the often broad posteriors that arise from partial orbits. The core computations of \texttt{ethraid} are implemented in Cython for speed. We validate \texttt{ethraid}'s performance by using it to constrain the masses and separations of the planetary companions to HD 117207 and TOI-1694. We designed \texttt{ethraid} to be both fast and simple, and to give broad, "quick look" constraints on companion parameters using minimal data. \texttt{ethraid} is pip installable and available on Github\footnote{https://github.com/jvanzand/ethraid}.

\keywords{Exoplanets, Bayesian statistics, Radial velocity, Astrometry, Gaia, Direct imaging}
\end{abstract}

\section{Introduction}

Fitting the orbits of exoplanets is one of the best tools for uncovering their origins, evolutionary history, and relationship to neighboring planets. Keplerian fits to RV observations produced the first detections of hot Jupiters orbiting Sun-like stars \citep{MayorQueloz1995, Butler1997, Cochran1997}, and higher-precision measurements have facilitated mass determinations of smaller planets \citep{Pinamonti2018, Luque2019}.

Planets with long periods, from years to decades, present multiple observational challenges. In particular, their orbits require consistent and prolonged monitoring to characterize. Long-term RV surveys (see, e.g., \citealt{Rosenthal2021}) confront this problem directly by compiling observational baselines over multiple decades. Though fruitful, these efforts are costly in terms of both human time and telescope time.

Furthermore, these surveys are observationally inefficient in two ways. First, the precise mass and orbital parameters of each companion are often not of immediate interest. For example, studies of giant planet occurrence rates \citep{Fulton2021} or the correlation between giants and other planet classes \citep{Bryan2019, Rosenthal2022} are principally concerned with differentiating between giant planets, brown dwarfs, and stars, which may be accomplished with less than a full orbit. Although precise constraints may be useful, some science cases benefit from greater statistical breadth over precision measurements of specific orbital parameters. 

Second, once legacy surveys end, their final catalogues may still include targets with too little phase coverage to characterize \citep{Rosenthal2021}. Without further observation, these time series represent significant investments of telescope time with minimal scientific return. Both of these limitations motivate tools for extracting planetary information from partial orbits.

Many existing Bayesian orbit fitters are well-suited to performing precise fits using RVs (\texttt{radvel}, \citealt{radvel}; \textit{The Joker}, \citealt{Joker2017}), astrometry (\texttt{orbitize!},  \citealt{orbitize2020}; \texttt{OFTI}, \citealt{OFTI2017}), or both (\texttt{orvara}, \citealt{orvara}). Some of these codes are designed to be robust against specific orbit fitting challenges (e.g., non-Gaussian posteriors, low phase coverage, and high-dimensional parameter spaces). However, their performance tends to suffer as these limitations are taken to the extreme.

In this work, we present \texttt{ethraid}, an open source Python package designed specifically to constrain companion mass and separation given partial phase coverage. \texttt{ethraid} can simultaneously model summary statistics of up to three independent data types: linear/quadratic RV trends, astrometric trends from the \textit{Hipparcos-Gaia} Catalog of Accelerations \citep{Brandt2018, Brandt2021}, and direct imaging contrast curves. In the remainder of this paper, we refer to these summary statistics as the "data."

This paper takes the following structure. In Section \ref{sec:fitting_algo} we describe \texttt{ethraid}'s fitting algorithm, including parameter sampling, forward modeling and likelihood calculation, and marginalization to derive posteriors. Section \ref{sec:forward_model} gives the mathematical details of our forward model for each of RVs, astrometry, and direct imaging data. We discuss \texttt{ethraid}'s performance in Section \ref{sec:performance} and basic usage in Section \ref{sec:usage}. We test the strengths and weaknesses of this code in Section \ref{sec:validation}. Finally, we give a brief list of future improvements planned for \texttt{ethraid} in Section \ref{sec:future_improvements} before concluding in Section \ref{sec:conclusion}.

\section{\texttt{ethraid}'s Fitting Algorithm}
\label{sec:fitting_algo}

Given some RV, astrometric, or imaging data, \texttt{ethraid} samples a space of orbital models, assesses each model's probability of having produced the signal based on both its \textit{a priori} probability and the likelihood of the measured signal in light of the model, and uses the resulting posterior surface to calculate confidence intervals for the inferred companion's mass and semi-major axis. \texttt{ethraid}'s approach to this problem consists of three steps:

\begin{enumerate}
    \item Random sampling of orbital parameters from prior distributions
    \item Forward modeling and likelihood calculation
    \item Marginalization over orbital parameters
\end{enumerate}

\begin{figure*}
  \centering\includegraphics[scale=0.68]{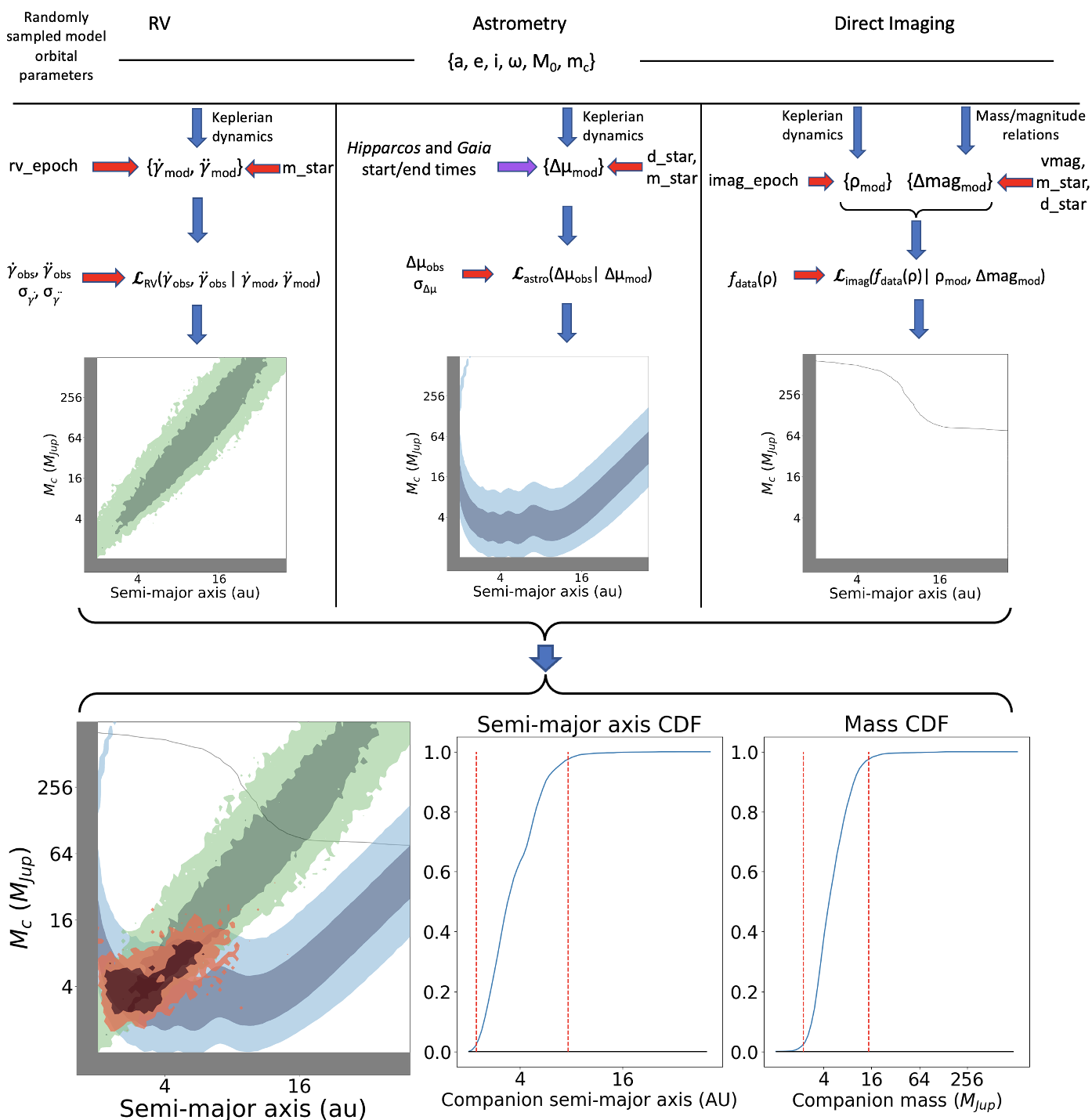}
 \centering\caption{\texttt{ethraid}'s fitting algorithm. The steps are the same for RVs (left), astrometry (center), and direct imaging (right): \texttt{ethraid} samples a set of model orbital parameters, forward models the data-specific parameters, evaluates the model likelihood, and finally marginalizes many such likelihoods to derive the posterior PDF. Arrows represent inputs used in a given step: blue arrows show inputs from the previous step, red arrows show inputs provided by the user, and purple arrows show inputs that are fixed in the code. The bottom row shows three of the plots available to the user through the \texttt{plot} command. \textbf{Left:} The 2D $m_c$-$a$ posteriors for each of the RV (green), astrometry (blue), and imaging (gray line) data sets, plus the combined posterior (red) conditioned on all three. The dark/light regions of each posterior show $68/95\%$ confidence intervals. For imaging, the gray contour marks the $95\%$ confidence boundary, with models below/above the contour ruled in/out by the data. \textbf{Right:} CDFs of the marginalized $a$ and $m_c$ posteriors. The vertical red lines indicate $95\%$ confidence intervals. \texttt{ethraid} also produces corresponding PDFs by default.}
  \label{fig:ethraid_flowchart}
\end{figure*}

These steps are summarized in Figure \ref{fig:ethraid_flowchart}.
Before detailing them below, we note an important assumption that \texttt{ethraid} makes, namely, that any measured signals were produced by a single bound companion. Consequently, RV and astrometric trends produced by multiple companions, stellar activity, or instrumental systematics will not be properly interpreted and may produce misleading results (see Section \ref{sec:validation}).

\subsection{Parameter Sampling from Priors}

The algorithm used to sample observations (in our case, model parameters) from a distribution is a key component of any Monte Carlo method. \texttt{ethraid} uses importance sampling \citep{Kloek1978}, an approach in which observations $\boldsymbol{\theta}$ are sampled according to an importance function $q(\boldsymbol{\theta})$. The result is that the histogram of a set of $N$ samples, $\{\boldsymbol{\theta}_n\} _{n=1}^{N}$, converges to $q$ in the limit of large $N$:

\begin{gather}
    \{\boldsymbol{\theta}_n\} _{n=1}^{N} \sim q(\boldsymbol{\theta}),
    \label{eq:sample_q}
\end{gather}

\noindent where "$\sim$" denotes "has the distribution of." However, because our goal is to calculate the posterior probability $p(\boldsymbol{\theta})$, we must weight each observation:

\begin{gather}
    \{\boldsymbol{\theta}_n \cdot w_n\}_{n=1}^{N} \sim p(\boldsymbol{\theta}),
    \label{eq:sample_p}
\end{gather}

\noindent where the weights $w_n$ are given by

\begin{gather}
    w_n = \frac{p(\boldsymbol{\theta}_n)}{q(\boldsymbol{\theta}_n)}.
    \label{eq:sample_w1}
\end{gather}

\noindent Following the earlier partial orbit fitters \texttt{OFTI} and \textit{The Joker}, we sample orbital elements directly from their respective priors, $\pi(\boldsymbol{\theta})$. This choice is motivated by the fact that partial orbit data is generally uninformative, resulting in prior-dominated posteriors. Conveniently, choosing $q = \pi(\boldsymbol{\theta})$ simplifies the weight of each model to the model likelihood:

\begin{gather}
    w_n = \frac{\mathcal{L}(\boldsymbol{\theta}_n)\pi(\boldsymbol{\theta}_n)}{\pi(\boldsymbol{\theta}_n)} = \mathcal{L}(\boldsymbol{\theta}_n).
    \label{eq:sample_w2}
\end{gather}

\noindent Thus, \texttt{ethraid} approximates the posterior surface $p(\boldsymbol{\theta})$ by creating a histogram of orbital models, each sampled from the appropriate prior distributions and weighted by its likelihood.

The first step of this procedure is to sample six orbital parameters according to their respective priors: semi-major axis $a$, eccentricity $e$, inclination $i$, argument of periastron of the companion (not the host star, as is sometimes the case in RV codes; see \citealt{HouseholderWeiss2023}) $\omega$, the mean anomaly at a reference epoch $M_0$, and companion mass $m_c$. As all data products we model are insensitive to the longitude of the ascending node ($\Omega$), we fix it to 0 arbitrarily. The prior PDFs for these parameters are:

\begin{itemize}
    \item[\sbt]{$\log\left ( \frac{a}{\text{1 AU}} \right ) \sim
    \mathcal{U}(a_{\text{min}}, a_{\text{max}})$}
    \item[\sbt]{$\log\left ( \frac{m_c}{\text{$1 \mj$}} \right ) \sim
    \mathcal{U}(m_{\text{c,min}}, m_{\text{c,max}})$}
    \item[\sbt]{$\cos(i) \sim \mathcal{U}(0, 1)$}
    \item[\sbt]{$\omega \sim \mathcal{U}(0, 2\pi)$}
    \item[\sbt]{$M_0 \sim \mathcal{U}(0, 2\pi)$}
    \item[\sbt]{$e \sim \mathcal{E}$$(a, m_c)$}
\end{itemize}

\noindent where $\mathcal{U}$ denotes a uniform distribution and $\mathcal{E}$ is a user-selected eccentricity distribution. Options for the latter include 

\begin{itemize}
    \item[--] \texttt{`zero'}: a constant distribution of $e=0$
    \item[--] \texttt{`uniform'}: a uniform distribution between 0 and 0.99
    \item[--] \texttt{`kipping'}: the piecewise beta distribution for short- and long-period exoplanets presented by \cite{Kipping2013}
    \item[--] \texttt{`piecewise'}: a combination of the \cite{Kipping2013} distribution for planets ($\frac{m_c}{\mj} \leq 13$), the beta distribution derived by \cite{Bowler2020} for brown dwarfs ($13 < \frac{m_c}{\mj} \leq 80$), and a uniform distribution between 0.1 and 0.8 for stars ($80 < \frac{m_c}{\mj}$) found by \cite{Raghavan2010}.
\end{itemize}

While the last of these is the most physically motivated in general, the others give users the option to include prior knowledge about a given system. For example, a dynamically cool system with multiple known planets on circular orbits may be unlikely to host an eccentric outer giant. In this case, a user might opt for the \texttt{`kipping'} distribution to favor lower-eccentricity models.

Our importance sampling strategy differs from MCMC sampling (e.g. \texttt{radvel}), where each model draw in a chain is dependent on the previous step. Drawing samples independently is advantageous for characterizing highly non-gaussian or multi-modal posterior surfaces, which arise often in partial orbit analysis. Our method is similar to the rejection sampling technique used by \texttt{OFTI} and \textit{The Joker}, which also draws models independently from a proposal distribution. However, rather than accepting or rejecting models based on their posterior probability, \texttt{ethraid} assigns each model a weight according to its likelihood. In practice, models with a high chance of rejection under rejection sampling are assigned vanishingly small weights under importance sampling, yielding similar performance.

\subsection{Forward Modeling and Likelihood Calculation}

After sampling a set of parameters, \texttt{ethraid} evaluates the model likelihood, that is, the probability of observing the measured data given the model under consideration. \texttt{ethraid} accomplishes this by generating synthetic RV, astrometric, and/or imaging data sets, and comparing them to the observed data. This procedure is repeated for a number of model orbits specified by the user (typically $10^6$--$10^8$). The likelihood calculations for each data type are detailed in Section \ref{sec:forward_model}.

In the case that multiple data sets are provided, \texttt{ethraid} records the likelihood of each data set conditioned on each model. Because the RVs, astrometry, and imaging are independent, \texttt{ethraid} calculates the combined likelihood of all provided data sets as the product of the likelihoods of each individual data set.

\subsection{Marginalization}
\texttt{ethraid} can generate visualizations of the 1D and 2D marginalized posteriors for $a$ and $m_c$. The marginalization step displays the desired posterior by summing the likelihoods of all models in the same interval of parameter space. For a 1D posterior, these are intervals in the desired parameter, for example, $\Delta m_{c,i}$. For a 2D posterior, they are interval pairs ($\Delta a_i$, $\Delta m_{c,i}$). 

\section{Forward model}
\label{sec:forward_model}

In this section we detail \texttt{ethraid}'s likelihood calculations for RV, astrometry, and imaging data, including the forward modeling procedure which produces an analytic counterpart to each measured quantity.

\subsection{RV Constraints}
\label{subsec:rv_constraints}

If a body on a long-period orbit is left unmodeled, it will present as a gradual increase or decrease in the RV residuals. For an orbital period much longer than the observing baseline, the variation is almost purely linear; for a period only a few times the baseline, there may also be significant quadratic curvature. The RV data therefore comprises the linear and quadratic coefficients to a second order polynomial fit to the RVs, denoted $\dot{\gamma}$ (m/s/day) and $\ddot{\gamma}$ (m/s/day$^2$), respectively.

For each set of sampled parameters, our goal is to forward model $\dot{\gamma}$ and $\ddot{\gamma}$, whose analytic forms are given by differentiating the stellar radial velocity $\gamma$:

\begin{gather}
    \begin{split}
        &\gamma = K \left [ e\cos \omega + \cos(\nu + \omega) \right ]\\
        &\dot{\gamma} = -K \left [ \dot{\nu} \sin(\nu + \omega) \right ]\\
        &\ddot{\gamma} = -K \left [ \dot{\nu}^2 \cos(\nu + \omega) + \ddot{\nu} \sin(\nu + \omega) \right ],
    \end{split}
    \label{eq:gamma_derivs}
\end{gather}

\noindent where $K$ is the RV semi-amplitude:

\begin{gather}
    K = \sqrt{\frac{G}{1-e^2}}\frac{m_c \sin i}{\sqrt{a(m_c+m_\star)}}
    \label{eq:K}
\end{gather}

\noindent and $\nu$ is the true anomaly, related to the eccentric anomaly $E$ by \citep{murray_dermott_2010}

\begin{gather}
    \nu = 2 \, \text{atan} \left [ \sqrt{\frac{1+e}{1-e}} \tan \frac{E}{2} \right ].
    \label{eq:nu_simple}
\end{gather}

\noindent The derivatives of $\nu$ are

\begin{gather}
    \begin{split}
        &\dot{\nu} = \frac{2 \pi \sqrt{1-e^2}}{P \left ( 1-e\cos E \right )^2}\\ \\
        &\ddot{\nu} = -\dot{\nu}^2 \frac{2e\sin E}{\sqrt{1-e^2}}.
    \end{split}
    \label{eq:nu_derivs}
\end{gather}

We obtain $E$ by numerically solving Kepler's equation:

\begin{gather}
    M = E - e\sin E,
    \label{eq:kepler}
\end{gather}

\noindent where $M$ is the mean anomaly, calculated by advancing the sampled initial mean anomaly $M_0$ some duration $t$ past a reference epoch via

\begin{gather}
    M(t) = M_0 + \frac{2\pi}{P}t,
    \label{eq:mean_anom}
\end{gather}

\noindent where $P$ is the orbital period calculated from Kepler's third law. We choose our reference epoch to be 1989.5 (see Section \ref{subsec:astrometry_constraints}).

We evaluate the model log-likelihood by calculating $\dot{\gamma}$ and $\ddot{\gamma}$ at the epoch of the RV observations and comparing them to their measured counterparts via

\begin{equation}
\begin{split}
    &\ln \left ( \mathcal{L}_{RV}(\boldsymbol{\theta}) \right )
    = -\frac{1}{2}\chi_{RV}^2 + C_{RV} \\ 
    &= -\frac{1}{2}\left(\frac{(\dot{\gamma}_{data} - \dot{\gamma})^2}
                         {\sigma_{\dot{\gamma}_{data}}^2}
                  + \frac{ (\ddot{\gamma}_{data} - \ddot{\gamma})^2}
                         {\sigma_{\ddot{\gamma}_{data}}^2}
                         \right) + C_{RV},
    \label{eq:likelihood_RV}
\end{split}
\end{equation}

\noindent where $\dot{\gamma}_{data}$ and $\ddot{\gamma}_{data}$ are the measured slope and curvature of the RV time series, $\sigma_{\dot{\gamma}_{data}}$ and $\sigma_{\ddot{\gamma}_{data}}$ are their respective uncertainties, and $C_{RV}$ is a constant.


From Equations \ref{eq:gamma_derivs} one may derive (see Appendix \ref{appendix:post_shapes}) that for small companion masses, surfaces of constant $\dot{\gamma}$ and $\ddot{\gamma}$ follow $m_c \propto a^2$ and $m_c \propto a^{7/2}$, respectively. Thus, for any sampled companion mass, there is a physical separation at which such a companion would produce the observed RV trend and curvature. In other words, RV constraints are generally consistent with a large range of ($a$, $m_c$) pairs. We seek to break this degeneracy by incorporating an independent data set: astrometry. 

\subsection{Astrometry Constraints}
\label{subsec:astrometry_constraints}

A star with no orbiting companions will have a constant proper motion. Meanwhile, a massive companion will induce a change in its host star's proper motion vector over some time interval, called the proper motion anomaly (PM$_a$; \citealt{Kervella2019}). Our goal is to model the magnitude of the PM$_a$ vector.

Our forward modeling procedure for astrometry is similar to the RV case, but is tailored to the data provided by the \textit{Hipparcos-Gaia} Catalog of Accelerations (HGCA; \citealt{Brandt2021}). This catalog aligned the reference frames of \textit{Hipparcos} (1989.85--1993.21, \citealt{HipparcosCatalog}) and \textit{Gaia} EDR3 (25 July 2014--28 May 2017; \citealt{GaiaEDR3}) to measure the accelerations of over 115,000 stars. We choose 1989.5 as the reference epoch $t_0$ for our RV, astrometric, and direct imaging models.

The HGCA reports three proper motions for each target: the epoch proper motions of \textit{Hipparcos} ($\vec{\mu}_H$) and \textit{Gaia} ($\vec{\mu}_G$) near 1991.25 and 2016.0, respectively, and the average proper motion given by the difference in position between these epochs divided by their ${\sim}25$-year baseline, $\vec{\mu}_{HG}$. The position-derived proper motion is generally the most precise proper motion measurement owing to the long baseline between the two missions, while $\vec{\mu}_G$ is usually the next-most precise. We therefore quantify PM$_a$ as $\Delta \vec{\mu}$, the difference between $\vec{\mu}_G$ and $\vec{\mu}_{HG}$, and model its norm:

\begin{gather}
    \Delta \mu = |\Delta \vec{\mu}| = |\vec{\mu}_G - \vec{\mu}_{HG}|.
\label{eq:dmu}
\end{gather}

\begin{figure}
  \centering\includegraphics[width=0.5\textwidth]{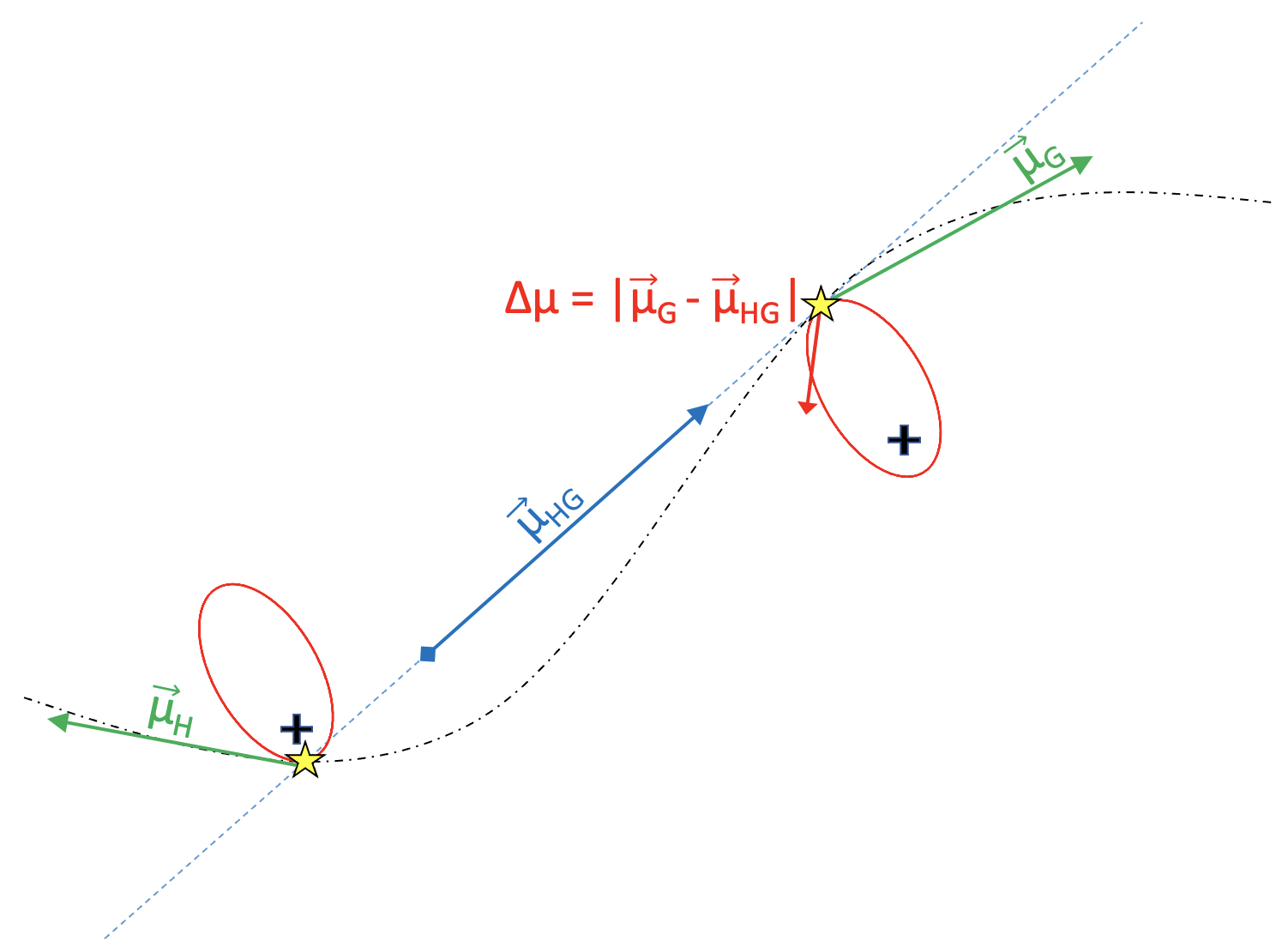}
 \centering\caption{Our model of astrometric proper motion anomaly. Red ellipses show the stellar orbit associated with a non-luminous bound companion (not shown). Green arrows show the average proper motion vectors fitted to the data from $Hipparcos$ (left) and $Gaia$ (right). Gold stars show the average positions during $Gaia$ and $Hipparcos$, and the blue arrow shows $\vec{\mu}_{HG}$, the average proper motion \textit{between} them, calculated as the difference in average positions divided by the difference between the mission midpoints. The proper motion anomaly $\Delta \mu$, shown in red, quantifies the difference between the average $Gaia$ proper motion and the position-derived proper motion. Figure adapted from \cite{Kervella2019}.}
  \label{fig:astro_diagram}
\end{figure}

\noindent We likewise obtain $\sigma_{\Delta \mu}$ by propagating the measurement uncertainties provided in the HGCA. The astrometric picture is summarized in Figure \ref{fig:astro_diagram}.

Our derivations of $\vec{\mu}_G$ and $\vec{\mu}_{HG}$ differ in one key respect from the procedure used to construct the HGCA. \cite{Brandt2021} obtained the best-fit positions and proper motions by recalibrating astrometric fits to the full $Hipparcos$ or $Gaia$ time series. Because simulating and fitting a model to a full time series would be computationally expensive, we instead calculate the analytic average of position and proper motion during each mission by integrating over the host star's orbit. We take these average quantities to approximate the stellar position and proper motion at the mission midpoints. To minimize positional uncertainty, Brandt evaluated the fitted positions at target-specific ``characteristic epochs" rather than the mission midpoints. This disparity introduces an error on the order of 0.01 mas/yr to our model, which we consider negligible because it is below both the median (0.047 mas/yr) and minimum (0.016 mas/yr) $\Delta \mu$ uncertainty in the HGCA.

Below we derive expressions for the stellar position and velocity due to a companion, and use them to model $\Delta \mu$. Calculating $\vec{\mu}_{HG}$ requires the average stellar position over each mission, while $\vec{\mu}_{G}$ requires the average stellar proper motion over only the $Gaia$ mission. 

In a two-body system composed of a host star and a bound companion, the star's position in the orbital plane relative to the system barycenter is (\citealt{murray_dermott_2010}, eq. 2.41)

\begin{equation}
\begin{aligned}
    X = -a_{\star}(\cos E - e)\\
    Y = -a_{\star}\sqrt{1-e^2} \sin E,
    \label{eq:xy_pos_astro}
\end{aligned}
\end{equation}

\noindent where $a_{\star} = a \frac{m_c}{m_{\star} + m_c}$ is the semi-major axis of the star's orbit and the positive $X$-axis is in the direction of the companion's periastron. We use uppercase letters to refer to coordinates in the orbital plane, and lowercase letters for the sky plane. The average position over a time interval [$t_1$, $t_2$] can be calculated analytically in the usual way: for the $X$-component, it is

\begin{gather}
    \langle X \rangle = \frac{1}{t_2 - t_1} \int^{t_2}_{t_1} X(t) \: dt,
\end{gather}

\noindent with a similar formula for $\langle Y \rangle$. We may reformulate this integral as:

\begin{gather}
    \langle X \rangle = \frac{1}{t_2 - t_1} \int^{E(t_2)}_{E(t_1)} X(E) \frac{dt}{dE} \, dE.
\end{gather}

\noindent We have $X(E)$ from Equations \ref{eq:xy_pos_astro} and $\frac{dt}{dE} = \frac{P(1-e \cos E)}{2\pi}$ can be obtained from Equation \ref{eq:kepler}. Carrying out the above integral and the corresponding one for $Y$ yields

\begin{equation}
\begin{split}
    &\langle X \rangle = \frac{-a_{\star}P}{2\pi(t_2 - t_1)} \Big [ (1+e^2)\sin E - \frac{e}{4} (6E + \sin(2E)) \Big ] ^{E(t_2)} _{E(t_1)} \\ \\ 
    &\langle Y \rangle = \frac{a_{\star} P \sqrt{1-e^2}}{2\pi(t_2 - t_1)} \Big [ \cos E - \frac{e}{2} \cos^2(E) \Big ] ^{E(t_2)} _{E(t_1)}.
\end{split}
\label{eq:avg_pos}
\end{equation}

To calculate average proper motion over the same time interval [$t_1$, $t_2$], we simply evaluate Equations \ref{eq:xy_pos_astro} at the interval boundaries and divide their difference by the elapsed time:

\begin{equation}
    \begin{aligned}
        \langle \dot{X} \rangle =  \frac{-1}{t_2 - t_1} \Big [ a_{\star}(\cos E - e) \Big ] ^{E(t_2)} _{E(t_1)} \\ \\
        \langle \dot{Y} \rangle =  \frac{-1}{t_2 - t_1} \Big [ a_{\star}\sqrt{1-e^2} \sin E \Big ]^{E(t_2)} _{E(t_1)}.
    \end{aligned}
\label{eq:avg_pm}
\end{equation}

\noindent Equations \ref{eq:avg_pos} and \ref{eq:avg_pm} allow us to define the average stellar position and proper motion relative to the system barycenter in the orbital plane:

\begin{equation}
    \begin{aligned}
        \vec{r}_{orb} = 
            \begin{bmatrix}
                \langle X \rangle \\
                \langle Y \rangle \\
                0
            \end{bmatrix}, \quad
        \vec{v}_{orb} = 
            \begin{bmatrix}
                \langle \dot{X} \rangle \\
                \langle \dot{Y} \rangle \\
                0
            \end{bmatrix}.
    \end{aligned}
    \label{eq:avg_vecs}
\end{equation}

To express these vectors in the observer frame, that is, with the $x$- and $y$-axes in the sky plane and the $z$-axis toward the observer, we apply the rotation matrix \textbf{R} \citep{murray_dermott_2010}, which is given in its full form in Appendix \ref{appendix:rot_matrix}.

\begin{equation}
    \begin{split}
        \vec{r}_{obs} = \textbf{R} \, \vec{r}_{orb} \\ \\
        \vec{v}_{obs} = \textbf{R} \, \vec{v}_{orb}
    \end{split}
    \label{eq:vec_rotation_astro}
\end{equation}

We project these 3-dimensional vectors onto the sky by ignoring the $z$-component, and convert to angular coordinates using the system's distance from Earth. The anomalous position $\vec{\rho}_{anom}$ and proper motion $\vec{\mu}_{anom}$ over either $Hipparcos$ or $Gaia$ are given by

\begin{equation}
    \begin{aligned}
        \vec{\rho}_{anom} 
        &=
            \begin{bmatrix}
                \alpha_{anom}*\\
                \delta_{anom}
            \end{bmatrix}
        = \frac{1}{d}
            \begin{bmatrix}
                r_{obs, x} \\
                r_{obs, y}
            \end{bmatrix} \\
        \vec{\mu}_{anom} 
        &= 
        \begin{bmatrix}
                \mu_{\alpha*} \\
                \mu_{\delta}
            \end{bmatrix}
        = \frac{1}{d}
            \begin{bmatrix}
                v_{obs, x} \\
                v_{obs, y}
            \end{bmatrix},
    \end{aligned}
    \label{eq:mu_and_rho}
\end{equation}

\noindent and the absolute position and proper motion at the mission midpoint $t_{mid}$ are

\begin{equation}
\begin{aligned}
        \vec{\rho}_{abs} 
        &=
            \begin{bmatrix}
                \alpha_{abs}* \\
                \delta_{abs}
            \end{bmatrix}
        \\
        &= 
            \begin{bmatrix}
                \alpha_0* + \mu_{0, \alpha*} \cdot (t_{mid} - t_0) + \alpha_{anom}* \\
                \delta_0  + \mu_{0, \delta} \cdot (t_{mid} - t_0) + \delta_{anom}
            \end{bmatrix} \\
    \vec{\mu}_{abs} &= \vec{\mu}_0 + \vec{\mu}_{anom},
    \label{eq:mu_and_rho_abs}
\end{aligned}
\end{equation}

\noindent where the subscript `0' refers to the barycentric position and proper motion at the reference epoch $t_0$ and $\alpha * \equiv \alpha \cos \delta$, where $\delta$ is the Dec at the epoch at which $\alpha$ is evaluated.\footnote{For concision, we misuse the $\alpha*$ notation in the case of $\alpha_{abs}*$, as it is composed of multiple terms evaluated at different epochs.}

We use the above results to estimate 1) $\{\alpha_{H, abs}*, \, \delta_{H, abs} \}$, the stellar RA/Dec at the $Hipparcos$ midpoint $t_{H, mid}$; 2) $\{\alpha_{G, abs}*, \, \delta_{G, abs} \}$, the stellar RA/Dec at the $Gaia$ midpoint $t_{G, mid}$; and 3) $\{\mu_{\alpha, G, abs *}, \, \mu_{\delta, G, abs} \}$, the stellar proper motions in the RA/Dec directions at $t_{G, mid}$. We provide here the procedure for calculating the RA component of the position-derived proper motion $\mu_{HG, abs}$. The calculation for the Dec component is analogous.

\begin{align}
\begin{aligned}
    & \mu_{HG, \alpha, abs *} = \frac{\alpha_{G,abs}* - \alpha_{H,abs}*}{t_{G, mid} - t_{H, mid}}
    \\
    \begin{split}
    & = \frac{1}{t_{G, mid} - t_{H, mid}}\\ 
    & \qquad \qquad \bigl[ \left[ \alpha_0*+\alpha_{G, anom}* + \mu_{0, \alpha*} \cdot \left( t_{G, mid} - t_0 \right)  \right] \\
            & \qquad \quad - \left[ (\alpha_0* + \alpha_{H, anom}*) + \mu_{0, \alpha*} \cdot \left( t_{H, mid} - t_0 \right) \right] \bigr]
    \end{split}
    \\
    \begin{split}
    & = \frac{\alpha_{G, anom}* - \alpha_{H, anom}* + \mu_{0, \alpha*} \cdot \left( t_{G, mid} - t_{H, mid} \right)}{t_{G, mid} - t_{H, mid}}
    \end{split}
    \\
    \begin{split}
    & = \frac{\alpha_{G, anom}* - \alpha_{H, anom}*}{t_{G, mid} - t_{H, mid}} + \mu_{0, \alpha*}
    \end{split}
    \\
    \begin{split}
    & = \mu_{HG, \alpha, anom*} + \mu_{0, \alpha*}.
    \end{split}
\end{aligned}
\end{align}

\noindent Combining $\vec{\mu}_{HG, abs}$ with $\vec{\mu}_{G, abs}$ from Equations \ref{eq:mu_and_rho_abs}, we have our final result:

\begin{gather}
   \Delta \mu = |\vec{\mu}_{G, abs} - \vec{\mu}_{HG, abs}| = |\vec{\mu}_{G, anom} - \vec{\mu}_{HG, anom}|.
\label{eq:dmu_final}
\end{gather}

The modeled $\Delta \mu$ is thus a function only of the anomalous proper motions, i.e., those induced by the companion. As in Section \ref{subsec:rv_constraints}, we calculate the model log-likelihood as

\begin{equation}
\begin{split}
    \ln \left ( \mathcal{L}_{astro}(\boldsymbol{\theta}) \right ) &= -\frac{1}{2} \chi_{astro}^2 + C_{astro} \\
    &= -\frac{1}{2}\left(\frac{({\Delta \mu}_{data} - \Delta \mu)^2}
                         {\sigma_{{\Delta \mu}_{data}}^2} \right) + C_{astro}
    \label{eq:likelihood_astro}
\end{split}
\end{equation}

\subsection{Direct Imaging Constraints}
\label{subsec:imaging_constraints}

RV and astrometric trends are often insufficient to constrain a companion's properties, either because one data set is not available (e.g., any star that is not in the original $Hipparcos$ catalog does not have an HGCA acceleration) or because the constraints they do provide rule out the same regions of parameter space, leaving the same $a$-$m_c$ degeneracy described in Section \ref{subsec:astrometry_constraints}. In these cases, we turn to direct imaging to place an upper limit on the companion mass.

Our data is the measured contrast curve, a table of angular separations and corresponding magnitude differences indicating the dimmest detectable companion at a given projected separation from the host star. For simplicity, we treat the contrast curve as a step function with no associated uncertainties. That is, at a given angular separation, any companion dimmer than the listed limit is completely undetectable, whereas any companion brighter than the limit would be detected in all cases. Our goal is to calculate a model companion's angular separation and magnitude difference to compare to the contrast curve. We begin with the angular separation.

\subsubsection{Angular Separation Calculation}

Similar to Equations \ref{eq:xy_pos_astro}, we calculate the host-companion separation via

\begin{equation}
\begin{aligned}
    x = a(\cos E - e)\\
    y = a\sqrt{1-e^2} \sin E,
    \label{eq:xy_pos_imag}
\end{aligned}
\end{equation}

\noindent where we now use the full semi-major axis $a$ instead of $a_{\star}$. Rotating the separation vector into the sky plane with the rotation matrix \textbf{R} and dividing by the distance to the system $d$, we obtain the projected angular separation $\rho$:

\begin{equation}
        \vec{\rho} = \frac{\textbf{R}}{d} \,             
            \begin{bmatrix}
                x \\
                y \\
                0
            \end{bmatrix}\\.
    \label{eq:vec_rotation_imag}
\end{equation}

\noindent This approach is fully consistent with Sections \ref{subsec:rv_constraints} and \ref{subsec:astrometry_constraints} in that it incorporates all orbital parameters into the modeled quantity.

The angular separation calculation above accounts for the fact that each model companion's projected separation is a function not only of $a$, but also of $e$, $i$, $\omega$, and $E$. However, this detailed procedure is expensive, requiring nearly $300\%$ and $70\%$ the time of the RV and astrometry calculations, respectively. We detail an approximate calculation of $\rho$ in Appendix \ref{appendix:approx_angsep_calculation} which significantly decreases run time at the cost of a slight reduction in accuracy.

\subsubsection{Contrast Calculation}
After modeling a companion's angular separation, we seek to convert its mass $m_c$ to a $\Delta$mag contrast in the appropriate band. We model contrast as dependent on $m_c$ and the stellar $V_{mag}$ only. We linearly interpolated magnitude-mass relations for both stellar (\citealt{Pecaut&Mamajek2013}, Table 5) and brown dwarf (\citealt{Baraffe03}, Table 4) companions in multiple bandpasses to define a function $F(m_c)$ which converts a model companion mass to $\Delta$mag in whichever of the following bands most closely matches the band of the given contrast curve: \{$V$, $R$, $I$, $J$, $H$, $K$, $L'$\}. \cite{Pecaut&Mamajek2013} do not include magnitude relations for the $K$ or $L'$ bands, though they do for $K_s$ and $W1$. We approximated that $K_s \approx K$ ($2.15 \, \mu m \approx 2.2 \, \mu m$) and $W1$ $\approx$ $L'$ ($3.37 \, \mu m \approx 3.78 \, \mu m$) to concatenate the two tables. The brown dwarf mass-magnitude relations of  \cite{Baraffe03} correspond to mature systems ($5$ Gyr).

We also performed a linear interpolation of the measured contrast curve, yielding another function $f_{data}(\rho)$ which converts measured angular separation to measured $\Delta$mag contrast. The imaging `data' is then the function $f_{data}$, together with the assumption that at any separation $\rho'$, no companion with $\Delta \text{mag} < f_{data}(\rho')$ was detected.

We therefore have that for a given model separation $\rho$ and contrast $\Delta \text{mag} = F(m_c)$, both derived from the initial model parameters $\boldsymbol{\theta}$, the log-likelihood that no companions were detected with $\Delta \text{mag} < f_{data}(\rho)$ is given by

\begin{equation}
\begin{split}
    \ln ( \mathcal{L}_{imag} & (\boldsymbol{\theta})) \\
    & =
     \left\{
        \begin{array}{ll}
        -\infty & \Delta \text{mag} < f_{data}(\rho) \\
        0 & \Delta \text{mag} \geq f_{data}(\rho)
        \end{array} + C_{imag}
    \right.. 
\end{split}
\label{eq:likelihood_imag}
\end{equation}

\begin{figure*}
  \centering\includegraphics[width=1.0\textwidth]{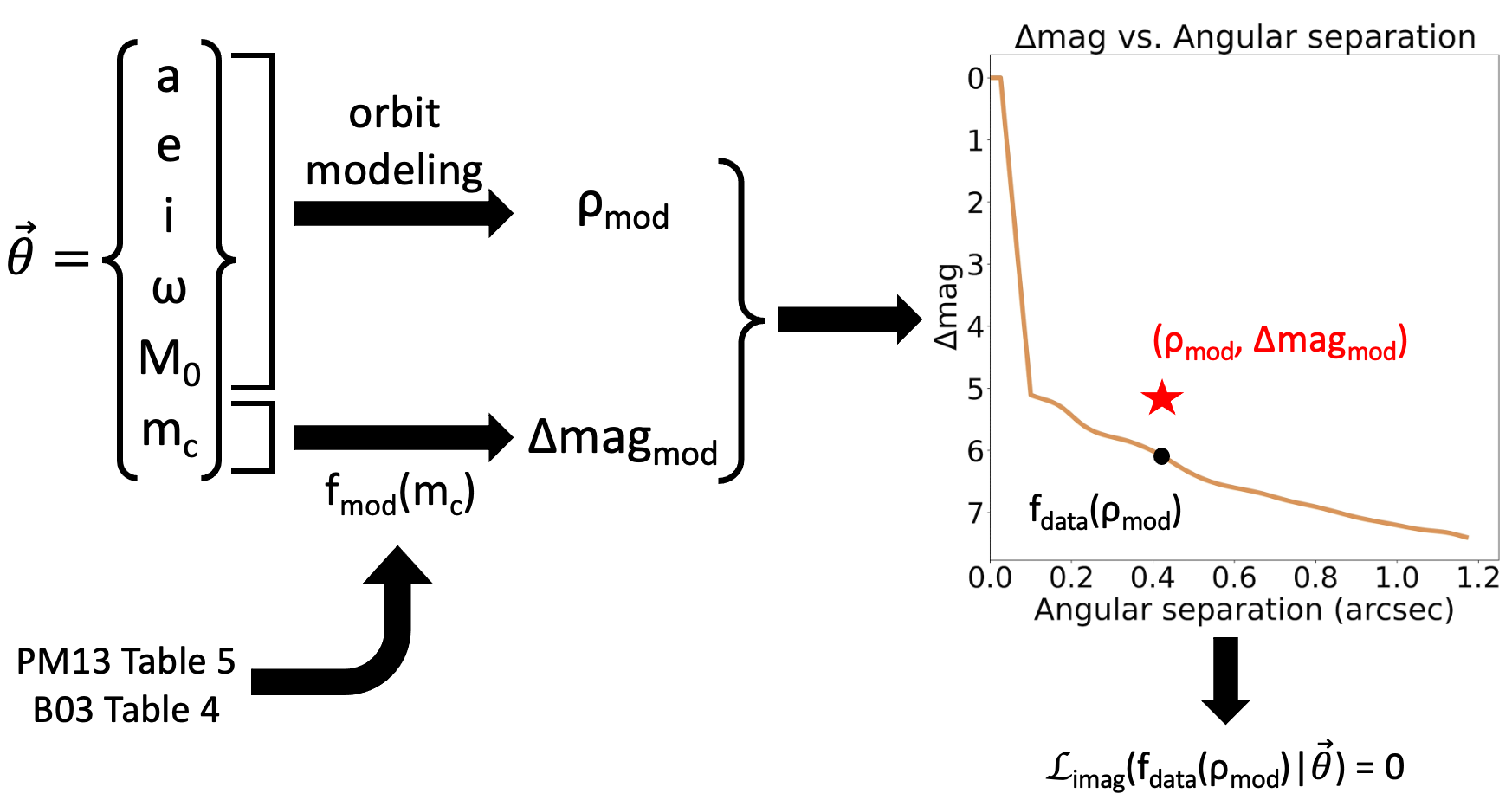}
 \centering\caption{\texttt{ethraid}'s likelihood calculation for direct imaging. Beginning with a set of sampled parameters, \texttt{ethraid} models angular separation (either \texttt{exactly} or \texttt{approximately}) using the orbital elements, and contrast using the companion mass, together with a linear interpolation function derived from Table 5 of \cite{Pecaut&Mamajek2013} and Table 4 of \cite{Baraffe03}, denoted by `PM13' and `B03,' respectively. \texttt{ethraid} then determines whether the model companion's brightness would have exceeded the minimum detectability threshold at the model separation in the real imaging data (gold line) and assigns a likelihood of 0 for detectable companions and 1 for non-detectable companions. The example model in the diagram falls above the threshold, so we assign it a likelihood of 0.}
  \label{fig:imag_diagram}
\end{figure*}

\noindent We visualize \texttt{ethraid}'s imaging likelihood calculation in Figure \ref{fig:imag_diagram}.

The independence of the RV, astrometric, and imaging measurements allows us to find the log-likelihood of all three data sets in light of a model orbit by simply summing Equations \ref{eq:likelihood_RV}, \ref{eq:likelihood_astro}, and \ref{eq:likelihood_imag}.

\section{Performance}
\label{sec:performance}

Four processes contribute meaningfully to the computational cost of an orbital fit. These include sampling orbital parameters from their prior PDFs and, for each sampled orbit, calculating the likelihood of each of the RV, astrometry, and/or direct imaging data sets. Therefore, the total time required to run an orbital fit is predetermined by the number of orbital models sampled (a value input by the user) and the number of data sets provided.

The time required to sample one model orbit and calculate the likelihood of all three data sets conditioned on that model is ${\sim}5.75 \, \mu$s using a single core of a 2.6 GHz Intel Core i7 processor. \texttt{ethraid} takes another 0.25 $\mu$s to multiply the three likelihoods to produce the combined likelihood, for a total of 6 $\mu$s per model. The average fraction of total run time required for each of these steps is: sampling - 6.8\%, RVs - 11.0\%, astrometry - 46.3\%, imaging - 31.8\%, combined likelihood - 4.1\%. If the approximate angular separation calculation is used for the imaging posterior instead of the detailed calculation (see Section \ref{subsec:imaging_constraints}), the share of time required by the imaging calculations drops to $<1\%$, reducing the overall run time by ${\sim} 30\%$.

\texttt{ethraid} implements the core calculations of Section \ref{sec:forward_model} in Cython, which provides a ${\sim} 5$x speedup over the standard Python equivalent functions, allowing users to produce informative posterior plots with $10^6-10^7$ sampled models on one core in $6-60$ seconds.

Two potential improvements to the current approach would substantially increase \texttt{ethraid}'s performance. First, the likelihood calculations are trivially parallelizable. Calculating the likelihoods for all data sets concurrently could decrease the total run time by $50\%$, and further parallelizing within data sets would give greater speedups. Second, the likelihood and sampled parameter lists tend to overload the working memory when the number of model orbits is large ($\gtrsim 10^8$), causing significant slowdowns. An alternative would be to store these arrays in temporary cached files, or to perform all necessary calculations associated with a given model or likelihood and then overwrite it immediately.

\section{Usage, fitting, saving, and plotting}
\label{sec:usage}

\texttt{ethraid} is installable through pip, and basic usage instructions are available on GitHub.

The typical use case of ethraid is for a system with an unexplained RV trend, though it may be run on systems with any combination of constraints from RVs, astrometry, or imaging.

The majority of parameters required to run an \texttt{ethraid} fit are stored in the corresponding system's \texttt{.py} configuration file. An \texttt{ethraid} configuration file contains all of the RV, astrometry, and direct imaging data needed to carry out the forward modeling algorithm laid out in Sections \ref{sec:fitting_algo} and \ref{sec:forward_model} (with the exception of the optional contrast curve for direct imaging constraints), as well as parameters for saving and plotting the calculated posteriors. We provide descriptions of these inputs in Table \ref{tab:config_params}.\footnote{\texttt{ethraid}'s configuration files are analogous to those used by \texttt{orvara}, and we model this table after Table 2 in \cite{orvara}.}

\begin{deluxetable*}{lccr}
\tablewidth{0pt}
\caption{Configuration file parameters}
\label{tab:config_params}

\tablehead{
            \colhead{Parameter Name} &
            \colhead{Data Type} &
            \colhead{Default Value} &
            \colhead{Description}
            }
\startdata
General & & & Parameters unrelated to any one data set, but necessary to perform fits\\
\hline
\texttt{num\_points} & Integer & \texttt{10}$^\texttt{6}$ & Number of model orbits to sample\\
\texttt{grid\_num} & Integer & [required] & Number of bins to divide parameter space into\\
\texttt{min\_a} & Float & \texttt{1} & Lower bound of semi-major axis sampling range (AU)\\
\texttt{max\_a} & Float & \texttt{100} & Upper bound of semi-major axis sampling range (AU)\\
\texttt{min\_m} & Float & \texttt{1} & Lower bound of companion mass sampling range (\mj)\\
\texttt{max\_m} & Float & \texttt{1000} & Upper bound of companion mass sampling range (\mj)\\
\texttt{e\_dist} & String & \texttt{`piecewise'} & \makecell[r]{Prior distribution from which model eccentricity values are sampled;\\ options: \{\texttt{`zero'}, \texttt{`uniform'}, \texttt{`kipping'}, \texttt{`piecewise'}\}}\\
\hline
Stellar & & & Parameters of the host star or whole system\\
\hline
\texttt{star\_name} & String & [required] & Host star name to label files; need not match any catalog\\
\texttt{m\_star} & Float & [required] & Host star mass (\mj)\\
\texttt{d\_star} & Float & [required] & Earth-system distance (AU)\\
\hline
RVs & & & \\
\hline
\texttt{run\_rv} & Bool & [required] & \makecell[r]{Whether to use RV data;\\ if False, all other RV arguments are optional}\\
\texttt{gammadot} & Float & --- & Linear RV trend (m/s/day)\\
\texttt{gammadot\_err} & Float & --- & Error on \texttt{gammadot} (m/s/day)\\
\texttt{gammaddot} & Float & --- & Quadratic RV curvature (m/s/day/day); setting to \texttt{None} excludes curvature\\
\texttt{gammaddot\_err} & Float & --- & Error on \texttt{gammaddot} (m/s/day/day); setting to \texttt{None} excludes curvature\\
\texttt{rv\_epoch} & Float & --- & Epoch approximating the midpoint of the RV baseline (BJD)\\
\hline
Astrometry & & & \\
\hline
\texttt{run\_astro} & Bool & [required] & \makecell[r]{Whether to use astrometry data;\\ if False, all other astrometry arguments are optional}\\
\texttt{delta\_mu} & Float & --- & Magnitude of proper motion change (mas/yr)\\
\texttt{delta\_mu\_err} & Float & --- & Error on \texttt{delta\_mu} (mas/yr)\\
\texttt{hip\_id} & String & --- & Hipparcos identifier of host star; alternative to providing \texttt{delta\_mu} directly\\
\texttt{gaia\_id} & String & --- & Gaia DR3 identifier of host star; alternative to providing \texttt{delta\_mu} directly\\
\hline
Direct Imaging & & & \\
\hline
\texttt{run\_imag} & Bool & [required] & \makecell[r]{Whether to use imaging data;\\ if False, all other imaging arguments are optional}\\
\texttt{imag\_calc} & String & \texttt{`exact'} & \texttt{`exact'} or \texttt{`approx'}: calculate imaging posterior exactly or approximately\\
\texttt{vmag} & Float & --- & Host star $V$ magnitude\\
\texttt{imag\_wavelength} & Float & --- & Wavelength of imaging data ($\mu$m)\\
\texttt{contrast\_str} & String & --- & Path to .csv file containing contrast curve\\
\texttt{imag\_epoch} & Float & --- & Epoch at which imaging data was taken (BJD)\\
\hline
Save & & & \\
\hline
\texttt{save} & List & [\texttt{`proc'}] & \makecell[r]{Save processed arrays (\texttt{`proc'}; binned 2D posteriors) and/or\\ raw arrays (\texttt{`raw'}; unbinned 1D likelihoods)}\\
\texttt{out\_dir} & String & \texttt{`'} & Path to save directory\\
\hline
Plot & & & \\
\hline
\texttt{scatter\_plot} & List & --- & List of ($a$, $m_c$) pairs to plot 1 or more known companion positions on 2D posterior\\
\hline
\enddata
\end{deluxetable*}

\subsection{Command Line Interface}
\label{subsec:cli}

\texttt{ethraid} supports three different commands from the command line: \texttt{run}, \texttt{plot}, and \texttt{lims}. \texttt{run} samples the six-dimensional parameter space and calculates a posterior for each data set provided, plus an additional combined posterior conditioned on all provided data sets. Using the \texttt{save} parameter in the configuration file, the user has the option to save the `raw' unbinned posteriors, which they may reshape as desired, or the `processed' posteriors, which are smaller and load more quickly but cannot be reshaped. All arrays are saved as \texttt{hdf5}\footnote{https://www.h5py.org/} files in the specified output directory.

After running a fit, the user may access and plot the saved arrays using the \texttt{plot} command. \texttt{plot} loads a set of posterior arrays from the specified file path and optionally produces and saves three figures. The first includes two one-dimensional marginalized posterior PDFs side by side, one for $a$ and the other for $m_c$. The second is analogous to the first, but shows CDFs instead of PDFs. The third is a two-dimensional plot in $m_c$-$a$ space displaying the posteriors conditioned on each data set individually as well as the combined posterior. We show examples of the one-dimensional CDFs and two dimensional PDFs in in Figure \ref{fig:ethraid_flowchart}.

The \texttt{lims} command simply prints the 2$\sigma$ confidence intervals (i.e., 2.5th and 97.5th percentiles) of both the $m_c$ and $a$ posterior PDFs to the terminal. It is intended to give a quick quantitative summary of the orbit fit. Users may load the raw or processed posterior arrays directly if they wish to calculate more detailed statistics. We caution that the bare numerical results returned by \texttt{lims} will not necessarily give an accurate summary of the posterior distribution, particularly in cases where the posterior is under-sampled. We encourage users to use both the one- and two-dimensional posterior plots to guide their interpretation of the confidence intervals returned by this command. 

Finally, the \texttt{all} command runs \texttt{run}, \texttt{plot}, and \texttt{lims} sequentially. We give more information on the command line interface in Table \ref{tab:cli_params}.

\begin{deluxetable*}{lccr}
\tablewidth{0pt}
\caption{Command line interface parameters}
\label{tab:cli_params}

\tablehead{
            \colhead{Argument Name} &
            \colhead{Data Type} &
            \colhead{Default Value} &
            \colhead{Description}
            }
\startdata
General & & & \\
\hline
\texttt{-{}-version} & --- & --- & Version number of current \texttt{ethraid} installation\\
\hline
All & & & Arguments common to all commands\\
\hline
\texttt{-od/-{}-outdir} & String & \texttt{`'} & Directory to which probability arrays and posterior plots will be saved\\
\texttt{-v/-{}-verbose} & --- & False & Whether to print verbose output; no arguments needed.\\
\hline
\texttt{run} & & & Command to perform orbital fit\\
\hline
\texttt{-cf/-{}-config} & String & [required] & Relative path of configuration file\\
\hline
\texttt{plot} & & & Command to plot fit results\\
\hline
\texttt{-cf/-{}-config} & String & [required] & Path to configuration file\\
\texttt{-rfp/-{}-read\_file\_path} & String & [required] & Path to saved \texttt{.h5} results file\\
\texttt{-t/-{}-type} & List & [required] & \makecell[r]{`1d' and/or `2d'; Whether to save 1d or 2d posterior plots \\ No brackets or quotation marks needed; separate multiple arguments by a space}\\
\hline
\texttt{lims} & & & Command to display $95\%$ $a$ and $M_p$ confidence intervals\\
\hline
\texttt{-rfp/-{}-read\_file\_path} & String & [required] & Path to saved \texttt{.h5} results file\\
\hline
\texttt{all} & & & Command to run the \texttt{run}, \texttt{plot}, and \texttt{lims} commands sequentially\\
\hline
\texttt{-cf/-{}-config} & String & [required] & Path to configuration file\\
\texttt{-rfp/-{}-read\_file\_path} & String & [required] & Path to saved \texttt{.h5} results file\\
\texttt{-t/-{}-type} & List & [required] & \makecell[r]{`1d' and/or `2d'; whether to save 1d or 2d posterior plots \\ No brackets or quotation marks needed; separate multiple arguments by a space}\\
\enddata
\end{deluxetable*}

\section{Validation}
\label{sec:validation}

We tested \texttt{ethraid} on four systems to judge its strengths and weaknesses in estimating companion parameters. We chose the first two systems, HD 117207 and TOI-1694, to illustrate \texttt{ethraid}'s capacity to characterize single companions with fractional orbital coverage. Both systems host RV-characterized giant planets with well-constrained orbital parameters. We find that \texttt{ethraid}'s parameter estimates are consistent with these `ground truth' values, regardless of the planet's phase during the interval of RV observation. We chose the other two systems, HD 114729 and HD 12661, to demonstrate \texttt{ethraid}'s failure modes. Both of these systems host two long-period giant companions, violating one of \texttt{ethraid}'s key assumptions that RV/astrometric signatures originate from a single companion. \texttt{ethraid}'s failure to accurately characterize the companions in either system illustrates the perils of applying this code blindly.

We designed our validation algorithm to mimic real applications of \texttt{ethraid}. For systems exhibiting periodic RV variability, we identified and isolated a subset of the RVs which we then subdivided into `slices' of equal duration. We chose the number of slices so that the RV variability was approximately linear/quadratic over each slice, emulating a real data set with limited coverage of the companion's phase. We used \texttt{radvel} to fit a second order polynomial to each slice and recorded the fitted trend and curvature terms. We then applied \texttt{ethraid} to each fitted trend/curvature pair and determined whether the results were consistent with the known companion parameters. For all fits, we used the \texttt{`piecewise'} eccentricity prior and simulated $10^8$ model orbits. We included astrometric trends from the \textit{Hipparcos-Gaia} Catalog of Accelerations (HGCA; \citealt{Brandt2021}) in our analysis when available.

\subsection{Case Study: HD 117207's 7.5-year super-Jupiter is identified using 1.8-year baselines}
\label{subsec:validation_117207}
HD 117207 is a chromospherically quiet ($\log R'_{HK}$=-5.06) G8 dwarf hosting a super-Jupiter companion, HD 117207 b, with minimum mass and separation initially reported as $2.06 \, \mj$ and $3.78$ AU \citep{Marcy2005}. \cite{Rosenthal2021}, using RVs spanning over 20 years, refined these values to $m_c \sin i = 1.87 \pm 0.075 \, \mj$ and $a = 3.744^{0.059}_{0.060}$ AU. 

\begin{figure*}
  \centering\includegraphics[scale=0.16]{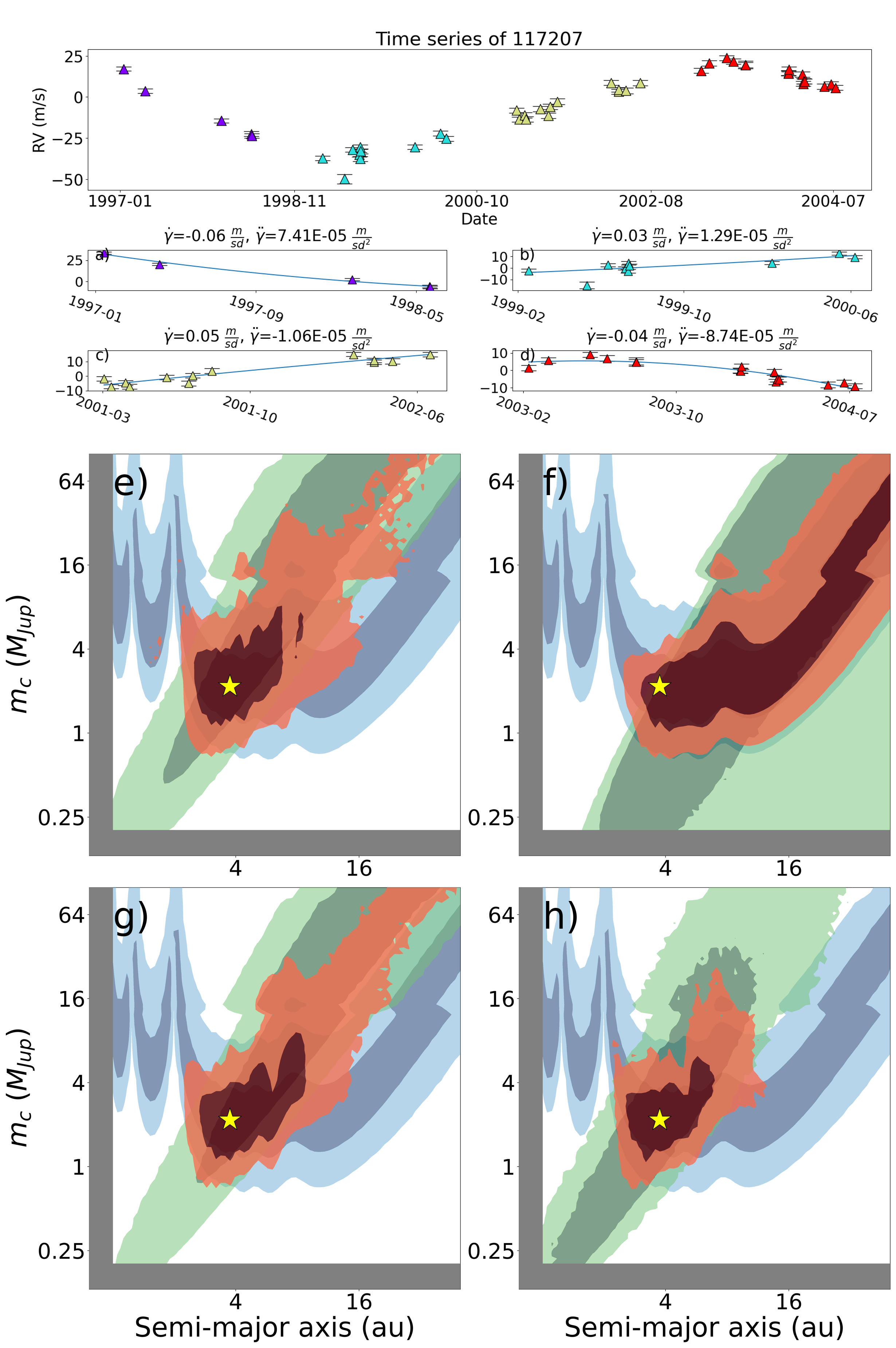}
 \centering\caption{The top panel shows a 7.5-year subset of HD 117207's RV time series, divided into four slices based on the epoch in which the RVs were measured. Panels \textbf{a)-d)} show the RVs from each slice, along with the linear/quadratic trend fit to those RVs using \texttt{radvel}. We show the measurement error of each RV with horizontal black lines. We used \texttt{ethraid} to model each trend/curvature pair, along with HGCA astrometry, in panels \textbf{e)-h)}, respectively. The gold star in each panel shows HD 117207 b's semi-major axis and $m_c \sin i$ values measured by long-baseline RV surveys: $a = 3.744$ AU, $m_c \sin i = 1.87$ \mj, where we have approximated a conversion from $m_c \sin i$ to $m_c$ by dividing by the median $\sin i$ value of 0.866. The overlap between \texttt{ethraid}'s predicted parameters and the true values in all four panels demonstrates \texttt{ethraid}'s reliability in estimating companion parameters over a range of orbital phases.}
  \label{fig:117207_long}
\end{figure*}

We identified a roughly 7.5-year section of the full time series and divided it into four slices. We used \texttt{radvel} \citep{radvel} to fit linear and quadratic trend terms to these slices. The results of this fitting procedure are shown in the top five panels of Figure \ref{fig:117207_long}.

We supplied the linear and quadratic trends from each slice as inputs to a partial orbit fit using \texttt{ethraid}. We also retrieved this star's astrometric trend of $0.13 \pm 0.03$ mas/yr from the HGCA and included it in our model. \texttt{ethraid} performed each joint RV/astrometric fit in about six minutes. We provide the posteriors of our partial orbit analysis of each slice in Figure \ref{fig:117207_long} and compare these results to the known parameters of HD 117207 b, shown in each panel as a gold star. In all four cases, the $68\%$ confidence interval derived by \texttt{ethraid} encompasses the true planet parameters.

We also validated our results for this system against \texttt{orvara}. We used the same HGCA astrometry and the raw RVs for each slice to emulate our \texttt{ethraid} fits. We ran 10 temperatures and 20 walkers over $10^5$ steps, which took about seven minutes (comparable to our \texttt{ethraid} runs). However, these fits did not converge and gave substantially broader mass and separation estimates than we obtained with \texttt{ethraid}. After increasing to $10^6$ steps, we obtained parameter estimates comparable to those from \texttt{ethraid} in about 80 minutes. We conclude that \texttt{ethraid}'s data compression approach allows it to match the performance of other robust fitters with minimal loss of information in the regime of highly limited orbital coverage and with only RVs and absolute astrometry.

\subsection{Case Study: TOI-1694's companion mass is bounded by direct imaging}
\label{subsec:validation_T001694}
TOI-1694 is an early K-dwarf with low chromospheric activity ($\log R'_{HK}$=-5.0) hosting a Jupiter analog with a minimum mass of $1.05 \pm 0.05 \, \mj$ at a separation of $0.98 \pm 0.01$ AU \citep{VanZandt2023}.

\begin{figure*}
  \centering\includegraphics[scale=0.14]{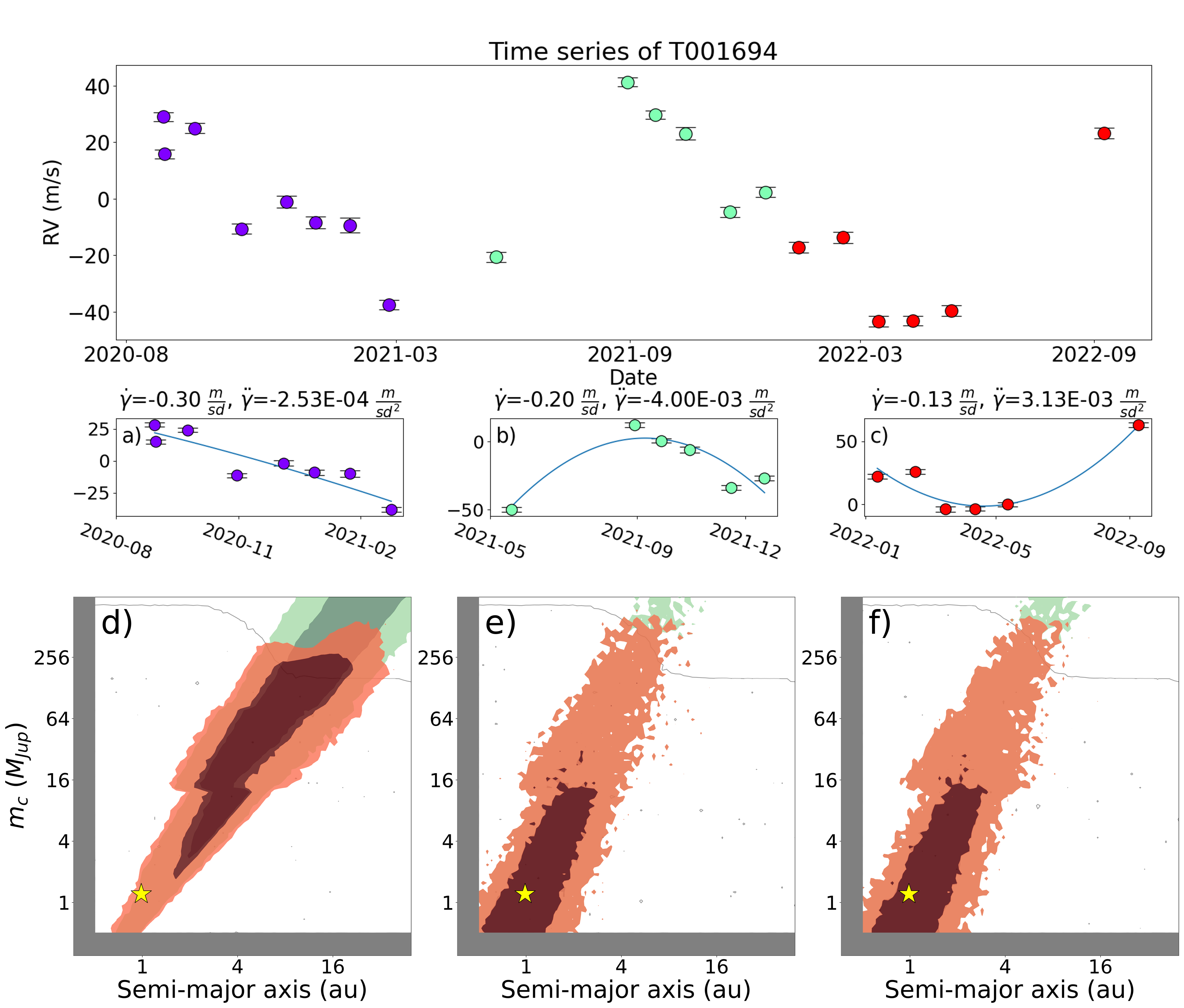}
 \centering\caption{
 Same as Figure \ref{fig:117207_long} for TOI-1694's two year baseline. The gray lines in panels \textbf{d)-f)} show posteriors derived from direct imaging. The gold stars show TOI-1694 b's median semi-major axis and adjusted $m_c \sin i$ values measured by \cite{VanZandt2023}: $a = 0.98$ AU, $m_c \sin i = 1.05$ \mj. Without astrometry, the trend and imaging data provide too little information to constrain the planet's mass and separation. However, they are able to rule out massive stars as well as the ($a, m_c$) pairs in the white regions. Note also that the posteriors in panels \textbf{e)} and \textbf{f)} favor shorter-period models due to the higher RV curvature associated with their trends.}
  \label{fig:T001694_long}
\end{figure*}

We divided the full two-year time series into three slices and fit linear and quadratic trend terms to them using \texttt{radvel} \citep{radvel}. The results of this fitting procedure are shown in the top four panels of Figure \ref{fig:T001694_long}.

We supplied the linear and quadratic trends from each slice as inputs to a partial orbit fit using \texttt{ethraid}. This star is not listed in the HGCA, and therefore has no available astrometric trend. However, we included direct imaging of this system \citep{Mistry2023} obtained in the $I$-band (832 nm) with the 'Alopeke speckle imager \citep{Scott2021} in our analysis. We provide the posteriors from our partial orbit analysis of each slice in Figure \ref{fig:T001694_long}. Although in this case the RV trend and imaging cannot disambiguate between planets, brown dwarfs, and low-mass stars, they are sufficient to rule out high-mass stars as well as planetary and brown dwarf companions beyond ${\sim} 10$ AU. Additionally, detection of curvature in some slices strongly favors models with shorter periods. The RV posteriors are consistent with TOI-1694 b's true parameters in all cases.

\subsection{Case Study: HD 114729's two companions conspire to produce misleading parameter estimates}
\label{subsec:validation_114729}

HD 114729 is an inactive ($\log R'_{HK}$=-5.02), metal-poor ([Fe/H]=-0.22) solar analog (G0 V) hosting a sub-Jupiter companion with a minimum mass and semi-major axis of $m_c \sin i = 0.892 \pm 0.053 \mj$ and $a = 2.094 \pm 0.022$ AU \citep{Rosenthal2021}. We refer to this planet as HD 114729 Ab. This system also hosts an M dwarf companion: \cite{Mugrauer2005} directly imaged HD 114729 and found a bound companion at a projected separation of 282 AU. Using the stellar cooling models of \cite{Baraffe1998}, as well as \cite{Butler2003}'s chromospherically derived age of 6 Gyr, \cite{Mugrauer2005} estimated HD 114729 B's mass to be $0.253 \, \msun$.

\begin{figure*}
  \centering\includegraphics[scale=0.16]{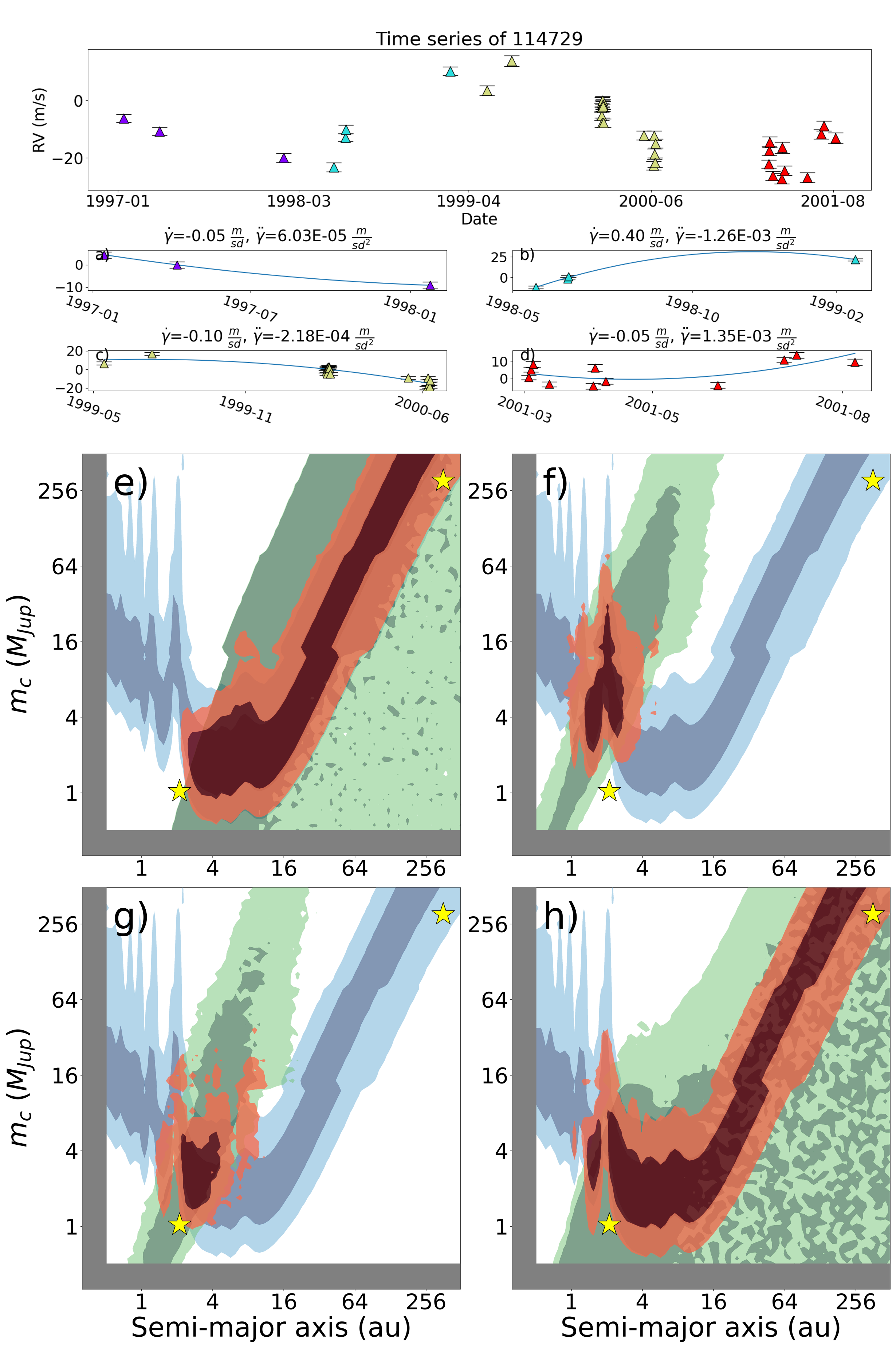}
 \centering\caption{
 Same as Figure \ref{fig:117207_long} for HD 114729. The gold star at the lower left of each panel shows the semi-major axis and adjusted $m_c \sin i$ of HD 114729 Ab  measured by long-baseline RV surveys: $2.094$ AU, $m_c \sin i = 0.892$ \mj. The upper right star shows the same for HD 114729 B, with a projected separation and mass of $282$ AU and $M = 0.253$ \msun. We estimated HD 114729 B's true separation by dividing the projected separation of 282 AU by 0.79, as described in Section \ref{subsec:imaging_constraints}. Agreement between the planet parameters and the RV posteriors indicates that \texttt{ethraid} is correctly recovering HD 114729 Ab's RV signature. Meanwhile, the consistency between HD 114729 B's parameters and the astrometric posterior suggests that HD 114729 B produced the astrometric trend. Note that in panels \textbf{a)} and \textbf{d)}, the fitted trend/curv values had precisions $< 2\sigma$, leading to broad RV posteriors in panels \textbf{e) and h)}.}
  \label{fig:114729_long}
\end{figure*}

We isolated a subset of the RV baseline covering approximately four years, divided it into four equal slices, and fit a second order polynomial to the RVs in each slice. The time series slices and polynomial fits are shown in the top of Figure \ref{fig:114729_long}.

We used \texttt{ethraid} to derive $m_c$-$a$ posteriors for each slice, incorporating both the fitted trend and curvature as well as HGCA astrometry ($0.11 \pm 0.03$ mas/yr), which is the same for all slices. In all cases, \texttt{ethraid}'s derived posterior distributions are discrepant with the true mass and semi-major axis of HD 114729 Ab at $> 2\sigma$ (Figure \ref{fig:114729_long}, bottom). This disagreement is driven by the astrometric data. The RV variability in the one-year slices is dominated by HD 114729 Ab because of its relatively short ${\sim}3$-year orbital period, and indeed the RV-only posteriors agree more closely with that planet's true parameters. Meanwhile, the astrometric variability reported in the HGCA is measured over 25 years, and thus includes a strong contribution from HD 114729 B's $4400$-year orbit. We also plot this companion's parameters in Figure \ref{fig:114729_long} to demonstrate its consistency with the astrometry data.

Our analysis of HD 114729 illustrates how \texttt{ethraid}'s fits may be misleading in the case of multiple companions. It is very valuable to provide direct imaging constraints whenever available to mitigate this failure mode.

\subsection{Case Study: HD 12661's two planets are misinterpreted as a single companion}
\label{subsec:validation_12661}

HD 12661 is an inactive ($\log R'_{HK}$=-5.06), metal-rich ([Fe/H]=0.293) G6 V star with two gas giant companions \citep{Fischer2001, Fischer2003}. HD 12661 b has a minimum mass of $m_c \sin i = 2.283^{+0.062}_{-0.063} \, \mj$ and a separation of $a = 0.824 \pm 0.011$ AU, while HD 12661 c has $m_c \sin i = 1.855 \pm 0.054 \, \mj$ and $a = 2.86^{+0.038}_{-0.039}$ AU \citep{Rosenthal2021}.

\begin{figure*}
  \centering\includegraphics[scale=0.13]{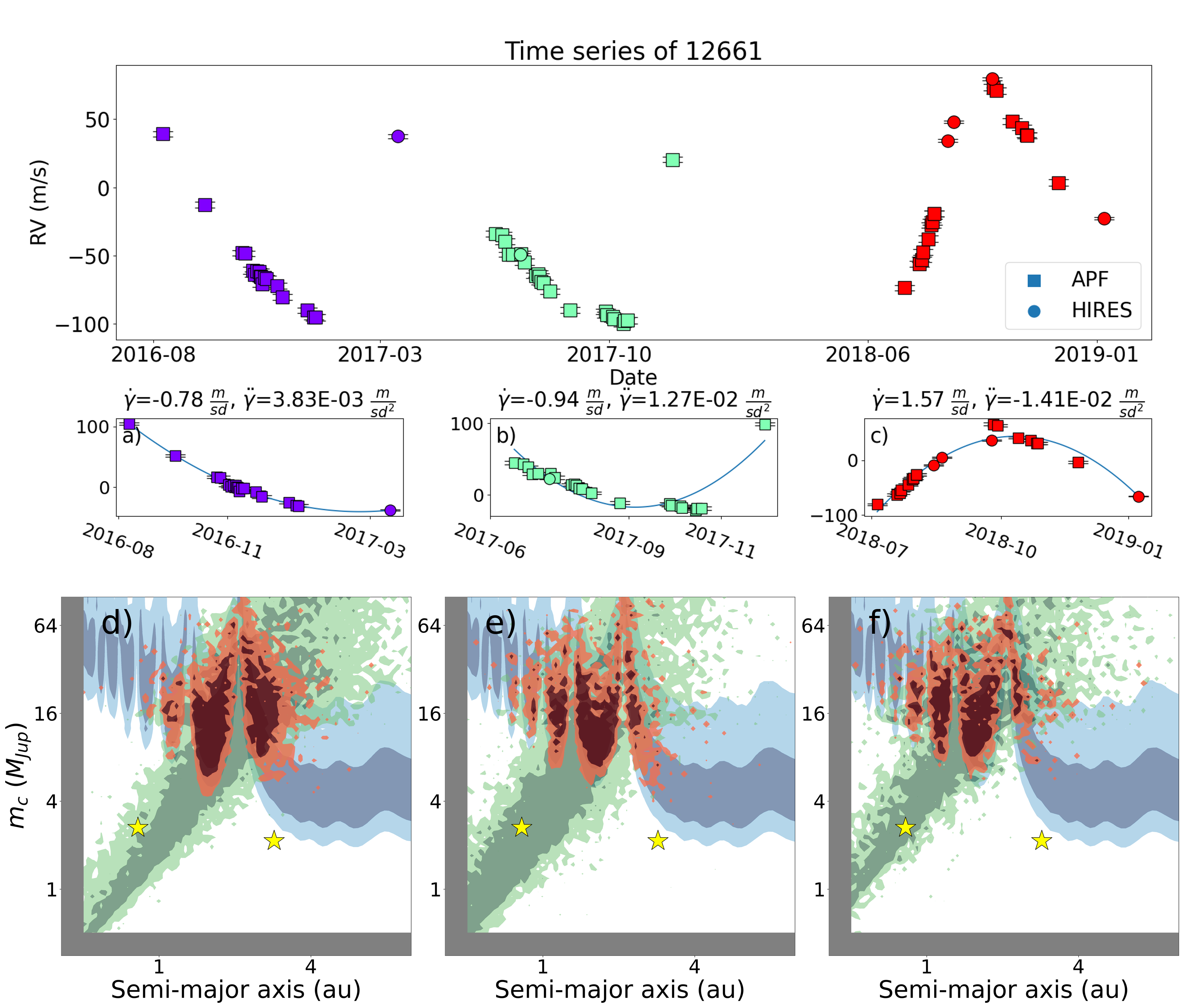}
 \centering\caption{
 Same as Figure \ref{fig:117207_long} for HD 12661. For this system, we include RVs from both the Keck/HIRES and APF/Levy instruments. The gold stars show the $a$ and adjusted $m_c \sin i$ values of HD 12661 b (left) and c (right). Planet b dominates the RV signature due to its short period, as shown by its consistency with the green RV posteriors in each slice. Meanwhile, planet c is responsible for the majority of the astrometric signal. The inconsistency between these planets' measured parameters and the combined posterior (red) shows that \texttt{ethraid} misinterpreted these separate contributions as originating from a single object of higher mass.}
  \label{fig:12661_long}
\end{figure*}

We chose a ${\sim} 2.5$-year subset of the RVs and divided it into three slices, each covering the same time interval. We fit linear and quadratic terms to each slice. The RV subset and slices are shown in the upper panels of Figure \ref{fig:12661_long}.

We used the trend and curvature derived for each slice in an \texttt{ethraid} partial orbit fit. We also included the astrometric trend of $0.23 \pm 0.05$ mas/yr. The orbit posteriors are shown in the lower panels of Figure \ref{fig:12661_long}, along with the the orbital parameters of HD 12661 b and c. \texttt{ethraid}'s RV posteriors are consistent with planet b's parameters, which is expected because with a larger mass than planet c and a period of 264 days, this planet dominates the RV variability over the 200-day slices. Furthermore, this consistency is not sensitive to the phase or curvature of the RV slices. The slices in Figure \ref{fig:12661_long} exhibit linear, quadratic, and higher order curvature, but are all nevertheless consistent with HD 12661 b's parameters in the lower panels. 

The astrometry posterior, on the other hand, is consistent with neither planet's parameters. We modeled the combined astrometric signature of both planets and found that planet c dominates this signal due to its wider separation. Furthermore, we found that if planet c's inclination is within 25 degrees of 0 (i.e., a "face-on" orbit), then the resulting underestimate of planet c's mass (a factor of $\gtrsim2.4$) may account for the disparity between this system's measured and expected astrometric signatures. A third, more distant companion could also explain this signal, though we did not explore this possibility in detail.

Like HD 114729, HD 12661 shows the limitations of \texttt{ethraid}'s approach to treating RV/astrometric trends. While planet b dominated the RV trend/curvature because of its short period, planet c had a much greater influence on the astrometric signal. \texttt{ethraid}'s assumption that both signals were due to one companion produced constraints inconsistent with either planet. We caution that unless the presence of multiple giant planets can be ruled out, \texttt{ethraid}'s single companion assumption may produce false parameter estimates.

\section{Future Improvements}
\label{sec:future_improvements}

Development of \texttt{ethraid} is ongoing, and we plan to implement a number of improvements in future versions, including:
\begin{enumerate}
    \item Parallelization of likelihood calculations.
    \item Dynamic likelihood storage to mitigate RAM overload.
    \item Analytical removal of the RV and astrometric signatures of known stellar companions, improving \texttt{ethraid}'s predictive power in systems like HD 114729 (Section \ref{subsec:validation_114729}).
    \item Improved companion mass prior to reflect the relative prevalences of planets, brown dwarfs, and low-mass stellar companions.
    \item Inclusion of brown dwarf cooling models for systems of ages other than 5 Gyr.
\end{enumerate}

We compiled the above improvements based primarily on our own use of \texttt{ethraid}, but we welcome contributions from the exoplanet community to make \texttt{ethraid} a useful and accessible tool.

\section{Conclusions}
\label{sec:conclusion}

In this work we presented \texttt{ethraid}, an open-source Python package for constraining the masses and separations of long-period companions using RVs, astrometry, and direct imaging. \texttt{ethraid} uses simple data inputs to calculate RV and astrometric posteriors, requiring only two values and their errors for the first and one value and its error for the second. Due to the limited information content of this data, \texttt{ethraid} is designed to broadly constrain companion parameters rather than to produce tight orbital fits, giving a `first look' at the nature of a companion and helping to disambiguate between multiple competing hypotheses (e.g., a close-in planet vs. a distant M dwarf). \texttt{ethraid} is \texttt{pip}-installable and may be downloaded from GitHub.

We demonstrated \texttt{ethraid}'s strengths and weaknesses by testing it on four systems hosting one or more known companions. HD 114729 \citep{Butler2003} hosts a giant planet and a stellar companion, and HD 12661 \citep{Fischer2001, Fischer2003} hosts two gas giant planets. In both cases, \texttt{ethraid}'s assumption of a single companion led to spurious parameter predictions. HD 117207 and TOI-1694 each host only one distant giant planet. \texttt{ethraid} consistently recovered HD 117207 b's parameters, and although a lack of astrometric data prevented a recovery of TOI-1694 b, our combined RV-direct imaging analysis ruled out high-mass stellar companions and high-separation ($\gtrsim 10$ AU) planets and brown dwarfs.

\texttt{ethraid} leverages up to three exoplanet detection techniques to constrain companion properties given minimal observational data. It is intended to help users evaluate the merit of potential companion models, as opposed to giving precise parameter constraints. We caution that due to \texttt{ethraid}'s implicit assumption that any input signals are caused by a single companion, applying it to multi-companion systems may give nonphysical results. Users may run \texttt{ethraid} on a system using one configuration file. We plan to maintain and improve \texttt{ethraid} so that it may be of use to the community.

\textit{Software:}
\texttt{radvel} \citep{radvel}

\section{Acknowledgments}

J.V.Z. acknowledges support from NASA FINESST Fellowship 80NSSC22K1606. J.V.Z. and E.A.P. acknowledge support from NASA XRP award 80NSSC21K0598.
We thank the anonymous reviewer for constructive input on our methodology, which significantly improved this manuscript. We also thank Tim Brandt for useful discussions regarding orbital fitting from limited data.

\bibliography{bib.bib}

\begin{thebibliography}{}
\expandafter\ifx\csname natexlab\endcsname\relax\def\natexlab#1{#1}\fi
\providecommand{\url}[1]{\href{#1}{#1}}
\providecommand{\dodoi}[1]{doi:~\href{http://doi.org/#1}{\nolinkurl{#1}}}
\providecommand{\doeprint}[1]{\href{http://ascl.net/#1}{\nolinkurl{http://ascl.net/#1}}}
\providecommand{\doarXiv}[1]{\href{https://arxiv.org/abs/#1}{\nolinkurl{https://arxiv.org/abs/#1}}}

\bibitem[{Hip(1997)}]{HipparcosCatalog}
 1997, ESA Special Publication, Vol. 1200, {The HIPPARCOS and TYCHO catalogues.
  Astrometric and photometric star catalogues derived from the ESA HIPPARCOS
  Space Astrometry Mission}

\bibitem[{{Baraffe} {et~al.}(1998){Baraffe}, {Chabrier}, {Allard}, \&
  {Hauschildt}}]{Baraffe1998}
{Baraffe}, I., {Chabrier}, G., {Allard}, F., \& {Hauschildt}, P.~H. 1998, \aap,
  337, 403, \dodoi{10.48550/arXiv.astro-ph/9805009}

\bibitem[{{Baraffe} {et~al.}(2003){Baraffe}, {Chabrier}, {Barman}, {Allard}, \&
  {Hauschildt}}]{Baraffe03}
{Baraffe}, I., {Chabrier}, G., {Barman}, T.~S., {Allard}, F., \& {Hauschildt},
  P.~H. 2003, \aap, 402, 701, \dodoi{10.1051/0004-6361:20030252}

\bibitem[{{Blunt} {et~al.}(2017){Blunt}, {Nielsen}, {De Rosa}, {Konopacky},
  {Ryan}, {Wang}, {Pueyo}, {Rameau}, {Marois}, {Marchis}, {Macintosh},
  {Graham}, {Duch{\^e}ne}, \& {Schneider}}]{OFTI2017}
{Blunt}, S., {Nielsen}, E.~L., {De Rosa}, R.~J., {et~al.} 2017, \aj, 153, 229,
  \dodoi{10.3847/1538-3881/aa6930}

\bibitem[{{Blunt} {et~al.}(2020){Blunt}, {Wang}, {Angelo}, {Ngo}, {Cody}, {De
  Rosa}, {Graham}, {Hirsch}, {Nagpal}, {Nielsen}, {Pearce}, {Rice}, \&
  {Tejada}}]{orbitize2020}
{Blunt}, S., {Wang}, J.~J., {Angelo}, I., {et~al.} 2020, \aj, 159, 89,
  \dodoi{10.3847/1538-3881/ab6663}

\bibitem[{{Bowler} {et~al.}(2020){Bowler}, {Blunt}, \& {Nielsen}}]{Bowler2020}
{Bowler}, B.~P., {Blunt}, S.~C., \& {Nielsen}, E.~L. 2020, \aj, 159, 63,
  \dodoi{10.3847/1538-3881/ab5b11}

\bibitem[{{Brandt}(2018)}]{Brandt2018}
{Brandt}, T.~D. 2018, \apjs, 239, 31, \dodoi{10.3847/1538-4365/aaec06}

\bibitem[{{Brandt}(2021)}]{Brandt2021}
---. 2021, \apjs, 254, 42, \dodoi{10.3847/1538-4365/abf93c}

\bibitem[{{Brandt} {et~al.}(2021){Brandt}, {Dupuy}, {Li}, {Brandt}, {Zeng},
  {Michalik}, {Bardalez Gagliuffi}, \& {Raposo-Pulido}}]{orvara}
{Brandt}, T.~D., {Dupuy}, T.~J., {Li}, Y., {et~al.} 2021, \aj, 162, 186,
  \dodoi{10.3847/1538-3881/ac042e}

\bibitem[{{Bryan} {et~al.}(2019){Bryan}, {Knutson}, {Lee}, {Fulton}, {Batygin},
  {Ngo}, \& {Meshkat}}]{Bryan2019}
{Bryan}, M.~L., {Knutson}, H.~A., {Lee}, E.~J., {et~al.} 2019, \aj, 157, 52,
  \dodoi{10.3847/1538-3881/aaf57f}

\bibitem[{{Butler} {et~al.}(2003){Butler}, {Marcy}, {Vogt}, {Fischer}, {Henry},
  {Laughlin}, \& {Wright}}]{Butler2003}
{Butler}, R.~P., {Marcy}, G.~W., {Vogt}, S.~S., {et~al.} 2003, \apj, 582, 455,
  \dodoi{10.1086/344570}

\bibitem[{{Butler} {et~al.}(1997){Butler}, {Marcy}, {Williams}, {Hauser}, \&
  {Shirts}}]{Butler1997}
{Butler}, R.~P., {Marcy}, G.~W., {Williams}, E., {Hauser}, H., \& {Shirts}, P.
  1997, \apjl, 474, L115, \dodoi{10.1086/310444}

\bibitem[{{Cochran} {et~al.}(1997){Cochran}, {Hatzes}, {Butler}, \&
  {Marcy}}]{Cochran1997}
{Cochran}, W.~D., {Hatzes}, A.~P., {Butler}, R.~P., \& {Marcy}, G.~W. 1997,
  \apj, 483, 457, \dodoi{10.1086/304245}

\bibitem[{ethraid(2024)}]{ethraidDOI}
ethraid. 2024, {ethraid v2.4.3: Characterize long-period companions with
  partial orbits.}, 2.4.3,  Zenodo, \dodoi{10.5281/zenodo.10841630}

\bibitem[{{Fischer} {et~al.}(2001){Fischer}, {Marcy}, {Butler}, {Vogt},
  {Frink}, \& {Apps}}]{Fischer2001}
{Fischer}, D.~A., {Marcy}, G.~W., {Butler}, R.~P., {et~al.} 2001, \apj, 551,
  1107, \dodoi{10.1086/320224}

\bibitem[{{Fischer} {et~al.}(2003){Fischer}, {Marcy}, {Butler}, {Vogt},
  {Henry}, {Pourbaix}, {Walp}, {Misch}, \& {Wright}}]{Fischer2003}
---. 2003, \apj, 586, 1394, \dodoi{10.1086/367889}

\bibitem[{{Fulton} {et~al.}(2018){Fulton}, {Petigura}, {Blunt}, \&
  {Sinukoff}}]{radvel}
{Fulton}, B.~J., {Petigura}, E.~A., {Blunt}, S., \& {Sinukoff}, E. 2018, \pasp,
  130, 044504, \dodoi{10.1088/1538-3873/aaaaa8}

\bibitem[{{Fulton} {et~al.}(2021){Fulton}, {Rosenthal}, {Hirsch}, {Isaacson},
  {Howard}, {Dedrick}, {Sherstyuk}, {Blunt}, {Petigura}, {Knutson}, {Behmard},
  {Chontos}, {Crepp}, {Crossfield}, {Dalba}, {Fischer}, {Henry}, {Kane},
  {Kosiarek}, {Marcy}, {Rubenzahl}, {Weiss}, \& {Wright}}]{Fulton2021}
{Fulton}, B.~J., {Rosenthal}, L.~J., {Hirsch}, L.~A., {et~al.} 2021, arXiv
  e-prints, arXiv:2105.11584.
\newblock \doarXiv{2105.11584}

\bibitem[{{Householder} \& {Weiss}(2022)}]{HouseholderWeiss2023}
{Householder}, A., \& {Weiss}, L. 2022, arXiv e-prints, arXiv:2212.06966,
  \dodoi{10.48550/arXiv.2212.06966}

\bibitem[{{Kervella} {et~al.}(2019){Kervella}, {Arenou}, {Mignard}, \&
  {Th{\'e}venin}}]{Kervella2019}
{Kervella}, P., {Arenou}, F., {Mignard}, F., \& {Th{\'e}venin}, F. 2019, \aap,
  623, A72, \dodoi{10.1051/0004-6361/201834371}

\bibitem[{{Kipping}(2013)}]{Kipping2013}
{Kipping}, D.~M. 2013, \mnras, 434, L51, \dodoi{10.1093/mnrasl/slt075}

\bibitem[{Kloek {et~al.}(1978)Kloek, Kloek, \& Van~Dijk}]{Kloek1978}
Kloek, T., Kloek, T., \& Van~Dijk, H. 1978, Econometrica, 46, 1,
  \dodoi{10.2307/1913641}

\bibitem[{{Lindegren} {et~al.}(2021){Lindegren}, {Klioner}, {Hern{\'a}ndez},
  {Bombrun}, {Ramos-Lerate}, {Steidelm{\"u}ller}, {Bastian}, {Biermann}, {de
  Torres}, {Gerlach}, {Geyer}, {Hilger}, {Hobbs}, {Lammers}, {McMillan},
  {Stephenson}, {Casta{\~n}eda}, {Davidson}, {Fabricius}, {Gracia-Abril},
  {Portell}, {Rowell}, {Teyssier}, {Torra}, {Bartolom{\'e}}, {Clotet},
  {Garralda}, {Gonz{\'a}lez-Vidal}, {Torra}, {Abbas}, {Altmann}, {Anglada
  Varela}, {Balaguer-N{\'u}{\~n}ez}, {Balog}, {Barache}, {Becciani}, {Bernet},
  {Bertone}, {Bianchi}, {Bouquillon}, {Brown}, {Bucciarelli}, {Busonero},
  {Butkevich}, {Buzzi}, {Cancelliere}, {Carlucci}, {Charlot}, {Cioni},
  {Crosta}, {Crowley}, {del Peloso}, {del Pozo}, {Drimmel}, {Esquej}, {Fienga},
  {Fraile}, {Gai}, {Garcia-Reinaldos}, {Guerra}, {Hambly}, {Hauser},
  {Jan{\ss}en}, {Jordan}, {Kostrzewa-Rutkowska}, {Lattanzi}, {Liao}, {Licata},
  {Lister}, {L{\"o}ffler}, {Marchant}, {Masip}, {Mignard}, {Mints}, {Molina},
  {Mora}, {Morbidelli}, {Murphy}, {Pagani}, {Panuzzo}, {Pe{\~n}alosa Esteller},
  {Poggio}, {Re Fiorentin}, {Riva}, {Sagrist{\`a} Sell{\'e}s}, {Sanchez
  Gimenez}, {Sarasso}, {Sciacca}, {Siddiqui}, {Smart}, {Souami}, {Spagna},
  {Steele}, {Taris}, {Utrilla}, {van Reeven}, \& {Vecchiato}}]{GaiaEDR3}
{Lindegren}, L., {Klioner}, S.~A., {Hern{\'a}ndez}, J., {et~al.} 2021, \aap,
  649, A2, \dodoi{10.1051/0004-6361/202039709}

\bibitem[{{Luque} {et~al.}(2019){Luque}, {Pall{\'e}}, {Kossakowski},
  {Dreizler}, {Kemmer}, {Espinoza}, {Burt}, {Anglada-Escud{\'e}}, {B{\'e}jar},
  {Caballero}, {Collins}, {Collins}, {Cort{\'e}s-Contreras},
  {D{\'\i}ez-Alonso}, {Feng}, {Hatzes}, {Hellier}, {Henning}, {Jeffers},
  {Kaltenegger}, {K{\"u}rster}, {Madden}, {Molaverdikhani}, {Montes}, {Narita},
  {Nowak}, {Ofir}, {Oshagh}, {Parviainen}, {Quirrenbach}, {Reffert}, {Reiners},
  {Rodr{\'\i}guez-L{\'o}pez}, {Schlecker}, {Stock}, {Trifonov}, {Winn},
  {Zapatero Osorio}, {Zechmeister}, {Amado}, {Anderson}, {Batalha}, {Bauer},
  {Bluhm}, {Burke}, {Butler}, {Caldwell}, {Chen}, {Crane}, {Dragomir},
  {Dressing}, {Dynes}, {Jenkins}, {Kaminski}, {Klahr}, {Kotani}, {Lafarga},
  {Latham}, {Lewin}, {McDermott}, {Monta{\~n}{\'e}s-Rodr{\'\i}guez}, {Morales},
  {Murgas}, {Nagel}, {Pedraz}, {Ribas}, {Ricker}, {Rowden}, {Seager},
  {Shectman}, {Tamura}, {Teske}, {Twicken}, {Vanderspeck}, {Wang}, \&
  {Wohler}}]{Luque2019}
{Luque}, R., {Pall{\'e}}, E., {Kossakowski}, D., {et~al.} 2019, \aap, 628, A39,
  \dodoi{10.1051/0004-6361/201935801}

\bibitem[{{Marcy} {et~al.}(2005){Marcy}, {Butler}, {Vogt}, {Fischer}, {Henry},
  {Laughlin}, {Wright}, \& {Johnson}}]{Marcy2005}
{Marcy}, G.~W., {Butler}, R.~P., {Vogt}, S.~S., {et~al.} 2005, \apj, 619, 570,
  \dodoi{10.1086/426384}

\bibitem[{{Mayor} \& {Queloz}(1995)}]{MayorQueloz1995}
{Mayor}, M., \& {Queloz}, D. 1995, \nat, 378, 355, \dodoi{10.1038/378355a0}

\bibitem[{{Mistry} {et~al.}(2023){Mistry}, {Pathak}, {Prasad}, {Lekkas},
  {Bhattarai}, {Gharat}, {Maity}, {Kumar}, {Collins}, {Schwarz}, {Mann},
  {Furlan}, {Howell}, {Ciardi}, {Bieryla}, {Matthews}, {Gonzales}, {Ziegler},
  {Crossfield}, {Giacalone}, {Tan}, {Evans}, {He{\l}miniak}, {Collins},
  {Narita}, {Fukui}, {Pozuelos}, {Dressing}, {Soubkiou}, {Benkhaldoun},
  {Schlieder}, {Suarez}, {Barkaoui}, {Palle}, {Murgas}, {Srdoc}, {Goliguzova},
  {Strakhov}, {Gnilka}, {Lester}, {Littlefield}, {Scott}, {Matson}, {Gillon},
  {Jehin}, {Timmermans}, {Ghachoui}, {Abe}, {Bendjoya}, {Guillot}, \&
  {Triaud}}]{Mistry2023}
{Mistry}, P., {Pathak}, K., {Prasad}, A., {et~al.} 2023, \aj, 166, 9,
  \dodoi{10.3847/1538-3881/acd548}

\bibitem[{{Mugrauer} {et~al.}(2005){Mugrauer}, {Neuh{\"a}user}, {Seifahrt},
  {Mazeh}, \& {Guenther}}]{Mugrauer2005}
{Mugrauer}, M., {Neuh{\"a}user}, R., {Seifahrt}, A., {Mazeh}, T., \&
  {Guenther}, E. 2005, \aap, 440, 1051, \dodoi{10.1051/0004-6361:20042297}

\bibitem[{Murray \& Dermott(2010)}]{murray_dermott_2010}
Murray, C.~D., \& Dermott, S.~F. 2010, Solar system dynamics (Cambridge Univ.
  Press)

\bibitem[{{Pecaut} \& {Mamajek}(2013)}]{Pecaut&Mamajek2013}
{Pecaut}, M.~J., \& {Mamajek}, E.~E. 2013, \apjs, 208, 9,
  \dodoi{10.1088/0067-0049/208/1/9}

\bibitem[{{Pinamonti} {et~al.}(2018){Pinamonti}, {Damasso}, {Marzari},
  {Sozzetti}, {Desidera}, {Maldonado}, {Scandariato}, {Affer}, {Lanza},
  {Bignamini}, {Bonomo}, {Borsa}, {Claudi}, {Cosentino}, {Giacobbe},
  {Gonz{\'a}lez-{\'A}lvarez}, {Gonz{\'a}lez Hern{\'a}ndez}, {Gratton}, {Leto},
  {Malavolta}, {Martinez Fiorenzano}, {Micela}, {Molinari}, {Pagano}, {Pedani},
  {Perger}, {Piotto}, {Rebolo}, {Ribas}, {Su{\'a}rez Mascare{\~n}o}, \&
  {Toledo-Padr{\'o}n}}]{Pinamonti2018}
{Pinamonti}, M., {Damasso}, M., {Marzari}, F., {et~al.} 2018, \aap, 617, A104,
  \dodoi{10.1051/0004-6361/201732535}

\bibitem[{{Price-Whelan} {et~al.}(2017){Price-Whelan}, {Hogg},
  {Foreman-Mackey}, \& {Rix}}]{Joker2017}
{Price-Whelan}, A.~M., {Hogg}, D.~W., {Foreman-Mackey}, D., \& {Rix}, H.-W.
  2017, \apj, 837, 20, \dodoi{10.3847/1538-4357/aa5e50}

\bibitem[{{Raghavan} {et~al.}(2010){Raghavan}, {McAlister}, {Henry}, {Latham},
  {Marcy}, {Mason}, {Gies}, {White}, \& {ten Brummelaar}}]{Raghavan2010}
{Raghavan}, D., {McAlister}, H.~A., {Henry}, T.~J., {et~al.} 2010, \apjs, 190,
  1, \dodoi{10.1088/0067-0049/190/1/1}

\bibitem[{{Rosenthal} {et~al.}(2021){Rosenthal}, {Fulton}, {Hirsch},
  {Isaacson}, {Howard}, {Dedrick}, {Sherstyuk}, {Blunt}, {Petigura}, {Knutson},
  {Behmard}, {Chontos}, {Crepp}, {Crossfield}, {Dalba}, {Fischer}, {Henry},
  {Kane}, {Kosiarek}, {Marcy}, {Rubenzahl}, {Weiss}, \&
  {Wright}}]{Rosenthal2021}
{Rosenthal}, L.~J., {Fulton}, B.~J., {Hirsch}, L.~A., {et~al.} 2021, \apjs,
  255, 8, \dodoi{10.3847/1538-4365/abe23c}

\bibitem[{{Rosenthal} {et~al.}(2022){Rosenthal}, {Knutson}, {Chachan}, {Dai},
  {Howard}, {Fulton}, {Chontos}, {Crepp}, {Dalba}, {Henry}, {Kane}, {Petigura},
  {Weiss}, \& {Wright}}]{Rosenthal2022}
{Rosenthal}, L.~J., {Knutson}, H.~A., {Chachan}, Y., {et~al.} 2022, \apjs, 262,
  1, \dodoi{10.3847/1538-4365/ac7230}

\bibitem[{{Scott} {et~al.}(2021){Scott}, {Howell}, {Gnilka}, {Stephens},
  {Salinas}, {Matson}, {Furlan}, {Horch}, {Everett}, {Ciardi}, {Mills}, \&
  {Quigley}}]{Scott2021}
{Scott}, N.~J., {Howell}, S.~B., {Gnilka}, C.~L., {et~al.} 2021, Frontiers in
  Astronomy and Space Sciences, 8, 138, \dodoi{10.3389/fspas.2021.716560}

\bibitem[{{Van Zandt} {et~al.}(2023){Van Zandt}, {Petigura}, {MacDougall},
  {Gilbert}, {Lubin}, {Barclay}, {Batalha}, {Crossfield}, {Dressing}, {Fulton},
  {Howard}, {Huber}, {Isaacson}, {Kane}, {Robertson}, {Roy}, {Weiss},
  {Behmard}, {Beard}, {Chontos}, {Dai}, {Dalba}, {Fetherolf}, {Giacalone},
  {Henze}, {Hill}, {Hirsch}, {Holcomb}, {Howell}, {Jenkins}, {Latham}, {Mayo},
  {Mireles}, {Mo{\v{c}}nik}, {Murphy}, {Pidhorodetska}, {Polanski}, {Ricker},
  {Rosenthal}, {Rubenzahl}, {Seager}, {Scarsdale}, {Turtelboom}, {Vanderspek},
  \& {Winn}}]{VanZandt2023}
{Van Zandt}, J., {Petigura}, E.~A., {MacDougall}, M., {et~al.} 2023, \aj, 165,
  60, \dodoi{10.3847/1538-3881/aca6ef}

\end{thebibliography}

\appendix

\section{Appendix: Approximate Angular Separation Calculation}
\label{appendix:approx_angsep_calculation}

In this appendix we present an alternative method to the one described in Section \ref{subsec:imaging_constraints} which makes use of two simplifying approximations to reduce computation time. First, for the imaging likelihood only, \texttt{ethraid} approximates that all models that fall into the same $a$-$m_c$ bin have the same value of $a$ and $m_c$, namely, the geometric mean of the lower and upper bin limits. This simplification allows \texttt{ethraid} to evaluate the imaging model likelihood once per two-dimensional \textit{bin} rather than once per \textit{model}, reducing the number of likelihood evaluations from ${\sim}10^6$--$10^8$ to $N^2$, where $N$ is the number of bins in both $a$ and $m_c$ space ($N=100$ by default).

Second, \texttt{ethraid} models angular separation as dependent only on $a$. In particular, it makes the following approximation:

\begin{gather}
    \rho \approx \frac{\langle p \rangle}{d} = \frac{Ca}{d},
    \label{eq:approx_proj_sep}
\end{gather}

\noindent where  $\langle p \rangle$ is the projected physical separation $p$ averaged over a companion's possible inclinations and orbital positions, and $C$ is a constant. If $C$ equals 1, Equation \ref{eq:approx_proj_sep} gives the angular separation for a circular, face-on orbit. We calculate the true value of $C$ by averaging $p$, which itself is given by

\begin{equation}
    \vec{p} \approx \textbf{R}
        \begin{bmatrix}
            a \cos M \\
            a \sin M \\
            0
        \end{bmatrix}
    =  a\begin{bmatrix}
            \cos M \\
            \sin M \cos i\\
            \sin M \sin i
        \end{bmatrix}
\end{equation}

\noindent where the approximate equality comes from fixing $e=0$ (and $\omega=0$ arbitrarily) for simplicity, so that true anomaly is equivalent to mean anomaly. Noting that the probability density function of $i$ on $\left [ 0, \pi/2 \right ]$ is $P(i)=\sin i$ and $P(M) = 1/2\pi$ on $\left [ 0, 2\pi \right ]$, the average value of $p$ is

\begin{gather}
    \langle p \rangle = \frac{a}{2\pi}\int_{0}^{2\pi} \int_{0}^{\pi/2} \sqrt{\cos^2 M + \sin^2 M \cos^2 i} \, \sin i \, \text{d}i \, \text{d}M.
    \label{eq:avg_proj_sep}
\end{gather}

Note that the vector norm (i.e., the square root term) includes only the components of $\vec{p}$ that are in the sky plane. Evaluating this integral numerically, we find $\langle p \rangle / a = C = \pi/4$, meaning that the sky-projected separation is approximately 0.79 times the true separation on average.

\begin{figure}[!h]
    \centering
    \includegraphics[width=.5\linewidth]{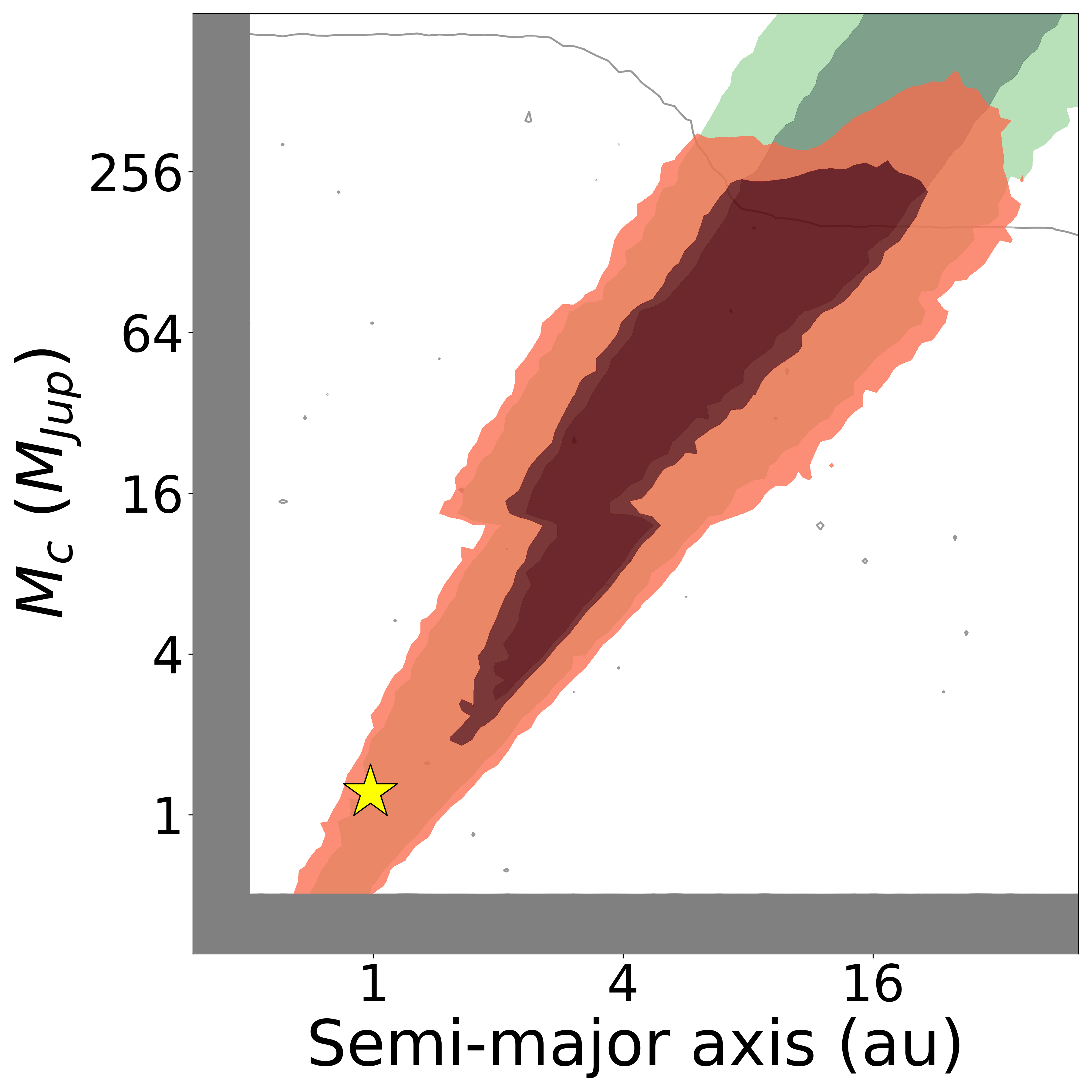}\hfill
    \includegraphics[width=.5\linewidth]{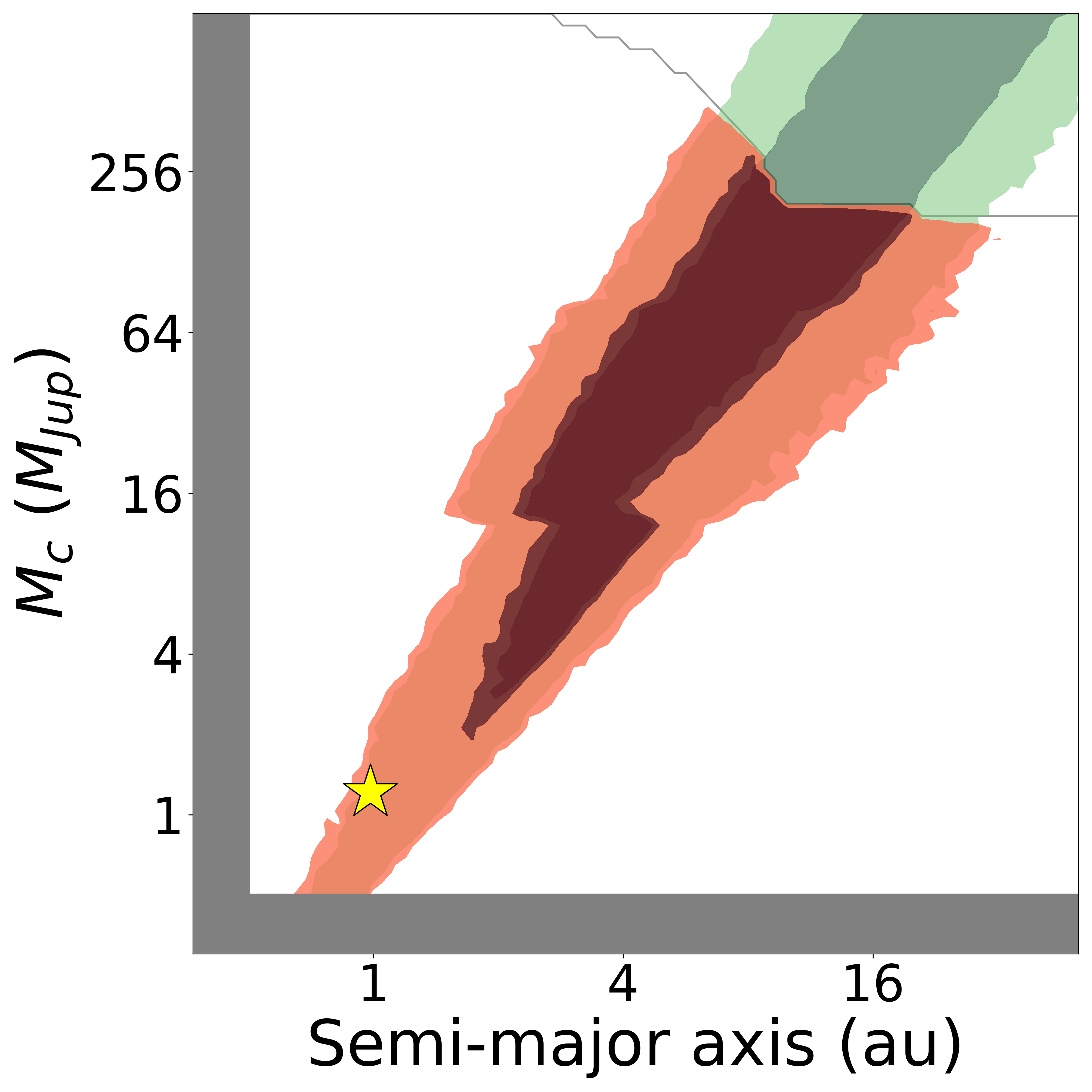}
    \caption{\textbf{Left:} Panel \textbf{d)} from Figure \ref{fig:T001694_long}, showing the companion models consistent with a subset of the RVs of TOI-1694, as well as the imaging data. Note that despite falling above the gray $95\%$ contour, some high-mass models are consistent with the imaging data because their orbital geometries result in small projected angular separations from the host star. \textbf{Right:} Parameter constraints for the same RV and imaging data, but using the approximate imaging calculation. In the approximate case, the imaging contour represents a hard bound, above which companions are ruled out with certainty. This causes the $1\sigma$ upper bound on mass (dark red) to be underestimated by $\sim30\%$ at 16 AU, while being roughly unaffected at 8 AU.}
    \label{fig:imag_exact_vs_approx}
\end{figure}

For systems where the RV and astrometric data rule out high-mass models, using the approximation described here may have a negligible effect. The disparity in parameter constraints is more pronounced when astrometry is not available, and imaging plays a more central role in ruling out stellar models. Figure \ref{fig:imag_exact_vs_approx} illustrates the effect of using this approximation for TOI-1694: some high-mass models that might in reality have gone undetected by the imaging data due to their orbital geometries are incorrectly ruled out. Using this approach to calculating angular separation is a suitable option for users who are comfortable incurring an error in their imaging constraints in exchange for cutting run time by a factor of ${\sim}1/3$.

\section{Appendix: Posterior Shapes}
\label{appendix:post_shapes}
The RV and astrometry posteriors derived in Section \ref{sec:forward_model} take on specific shapes, i.e., $m_c$-$a$ dependencies. Moreover, these dependencies vary as functions of $a$ and $m_c$ because different approximations hold in different regimes. We describe these dependencies here. Note that \texttt{ethraid} does not employ the following analytic descriptions during orbital fits; we provide them to aid the user in interpreting their results.

\subsection{RV Posterior}
We first reproduce Equations \ref{eq:gamma_derivs}--\ref{eq:nu_derivs}:

\begin{gather}
    \begin{split}
        &\gamma = K \left [ e\cos \omega + \cos(\nu + \omega) \right ] \\
        &\dot{\gamma} = -K \left [ \dot{\nu} \sin(\nu + \omega) \right ] \\
        &\ddot{\gamma} = -K \left [ \dot{\nu}^2 \cos(\nu + \omega) + \ddot{\nu} \sin(\nu + \omega) \right ]
    \end{split}
    \label{eq:gamma_derivs_appendix}
\end{gather}

\begin{gather}
    K = \sqrt{\frac{G}{1-e^2}}\frac{m_c \sin i}{\sqrt{a(m_c+m_\star)}}
    \label{eq:K_appendix}
\end{gather}

\begin{gather}
    \begin{split}
        &\dot{\nu} = \frac{2 \pi \sqrt{1-e^2}}{P \left ( 1-e\cos E \right )^2} \\ \\
        &\ddot{\nu} = -\dot{\nu}^2 \frac{2e\sin E}{\sqrt{1-e^2}}
    \end{split}
    \label{eq:nu_derivs_appendix}
\end{gather}

along with Kepler's third law,

\begin{gather}
    P^2 = \frac{4\pi^2 a^3}{G(m_\star + m_c)}.
    \label{eq:Kepler_third}
\end{gather}

From Equation \ref{eq:K_appendix}, we see that $K \propto \frac{m_c}{\sqrt{a(m_c + m_\star)}}$, while Equations \ref{eq:nu_derivs_appendix}--\ref{eq:Kepler_third} tell us that $\dot{\nu} \propto \sqrt{\frac{(m_c + m_\star)}{a^3}}$ and $\ddot{\nu} \propto \frac{(m_c + m_\star)}{a^3}$. Using these proportionalities in Equations \ref{eq:gamma_derivs_appendix}, we obtain

\begin{equation}
    \begin{aligned}
        \dot{\gamma} \propto \frac{m_c}{a^2}, \quad
        \ddot{\gamma} \propto \frac{m_c\sqrt{(m_c + m_\star)}}{a^{7/2}}.
    \end{aligned}
    \label{eq:gamma_proportionalities}
\end{equation}

\noindent \texttt{ethraid} explores different ($a$, $m_c$) combinations given fixed values of RV trend and curvature, so the posterior shapes are more clearly expressed by taking $\dot{\gamma}$ and $\ddot{\gamma}$ to be constant. In that case, we have

\begin{equation}
    \begin{aligned}
        m_c \propto a^2, \quad
        m_c\sqrt{(m_c + m_\star)} \propto a^{7/2}.
    \end{aligned}
    \label{eq:m_a_proportionalities}
\end{equation}

\begin{figure}[!h]
    \centering
    \includegraphics[width=.5\linewidth]{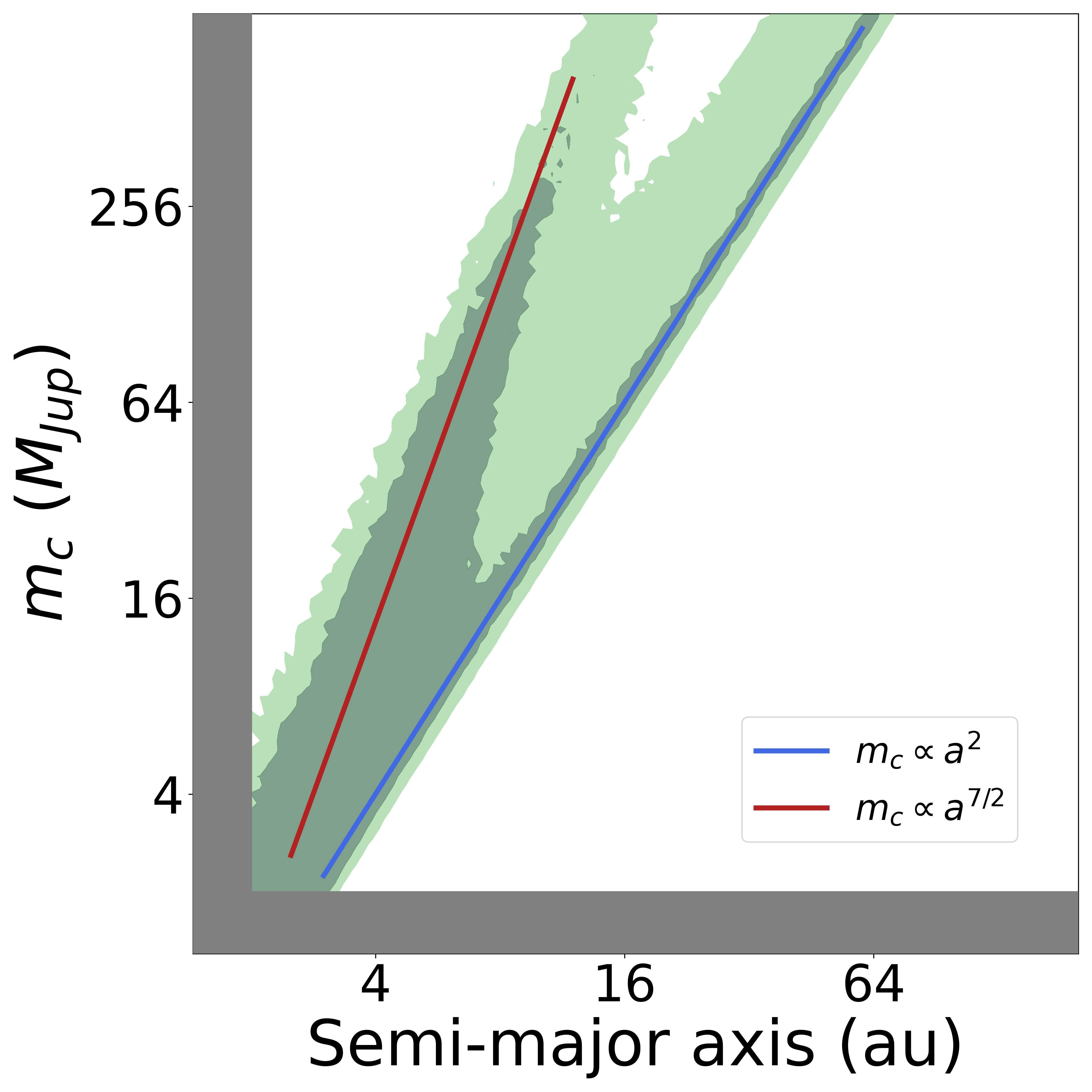}\hfill
    \includegraphics[width=.5\linewidth]{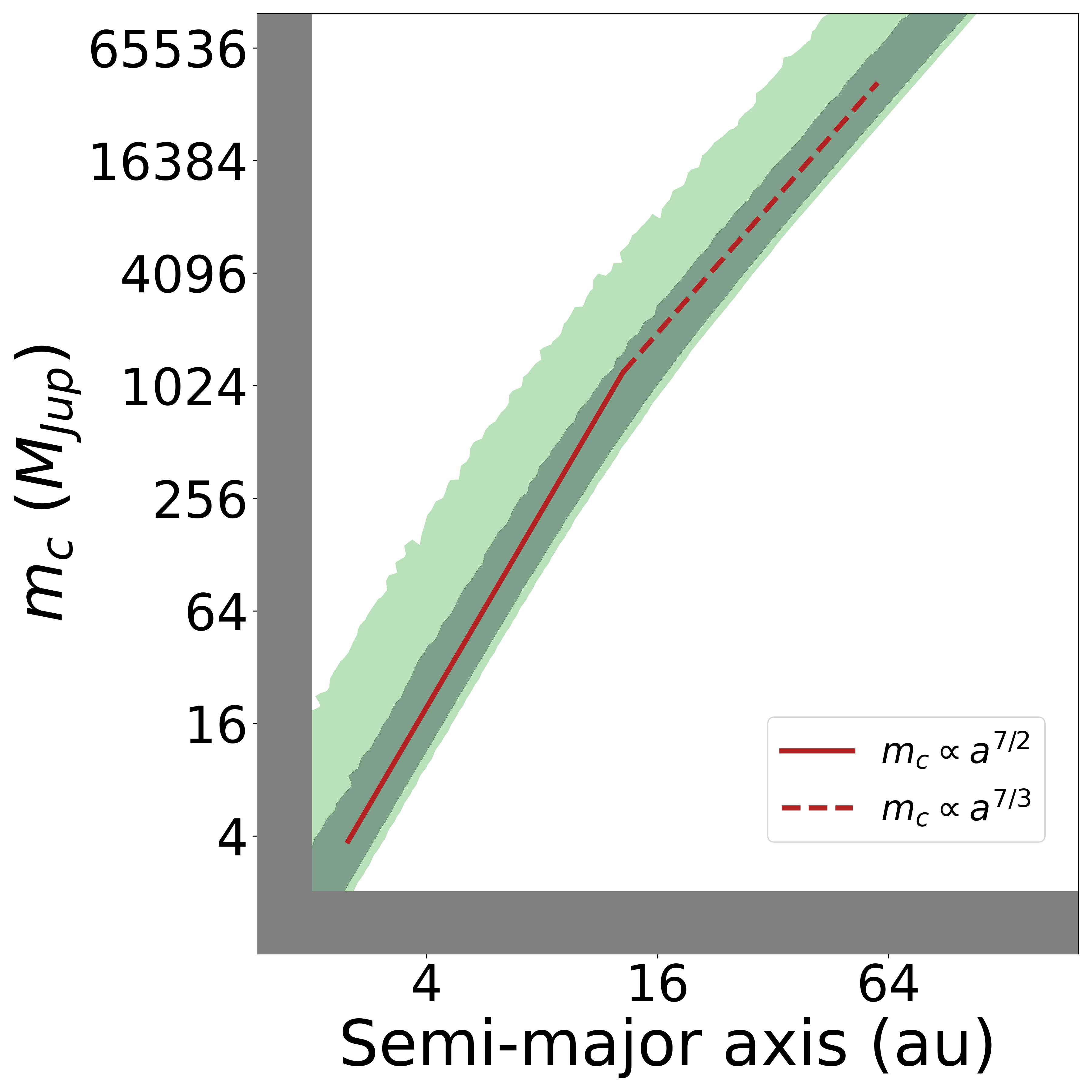}
    \caption{\textbf{Left:} An RV posterior calculated using both $\dot{\gamma}$ and $\ddot{\gamma}$. The blue line follows the relationship $m_c \propto a^{2}$, while the red line follows $m_c \propto a^{7/2}$. \textbf{Right:} An RV posterior calculated using $\ddot{\gamma}$ only, with a host star of 1 \msun. In the low-mass regime, the posterior follows $m_c \propto a^{7/2}$. However, the slope shifts to $m_c \propto a^{7/3}$ near 1 \msun.}
    \label{fig:rv_shapes}
\end{figure}

Because of $\ddot{\gamma}$'s dependence on $\sqrt{(m_c+m_\star)}$, the corresponding $m_c$-$a$ relation varies from $m_c \propto a^{7/2}$ to $m_c \propto a^{7/3}$ for $m_c << m_\star$ and $m_c \gtrsim m_\star$, respectively. In practice, most sampled companions have masses well below that of the host star, so $m_c \propto a^{7/2}$ prevails. Meanwhile, when both $\dot{\gamma}$ and $\ddot{\gamma}$ are provided, the RV posterior exhibits character of both relations in Equation \ref{eq:m_a_proportionalities}. Figure \ref{fig:rv_shapes} illustrates the interplay between $\dot{\gamma}$ and $\ddot{\gamma}$ in the RV posterior as well as $\ddot{\gamma}$ posterior's behavior at the transition between high and low companion masses.

\subsection{Astrometry Posterior}

\begin{figure}[!h]
    \centering
    \includegraphics[width=.9\linewidth]{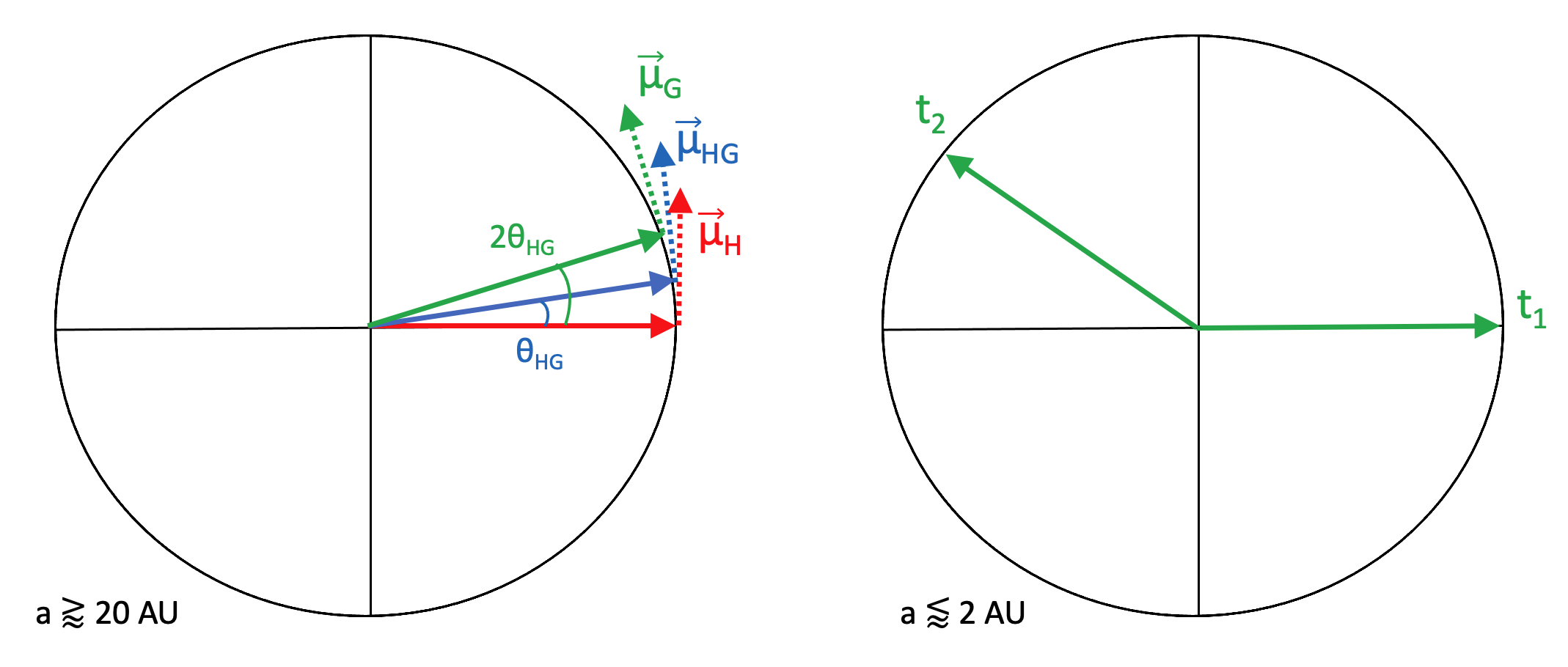}
    \caption{\textbf{Left:} A diagram of the circular, face-on orbit we use to understand the behavior of the astrometry posterior. The black circle shows the orbit of the host star about the system barycenter due to a companion at wide separation ($\gtrsim20$ AU). Solid and dotted arrows show the average stellar position and proper motion vectors, respectively, in each of the $Hipparcos$, $HG$, and $Gaia$ EDR3 epochs. $\theta_{HG}$ gives the position angle at the $HG$ epoch relative to the position angle during the $Hipparcos$ epoch. In the long-period regime, using the small angle approximation for $\theta_{HG}$ is appropriate. \textbf{Right:} A similar diagram for short periods ($\lesssim2$ AU). Arrows $t_1$ and $t_2$ show the position of the host star at the beginning and end of the $Gaia$ EDR3 data collection window. We chose $t_2$'s position  arbitrarily to illustrate that for short periods, the angle between the two position vectors is not small, but rather can take any value on [$0, 2\pi$]. We do not show the analogous vectors for $\vec{\mu}_{HG}$ because $T_{HG}$ is nearly nine times the duration of $T_G$; this larger denominator drives $\vec{\mu}_{HG}$ to nearly zero for short-period orbits.}
    \label{fig:astro_short_long_diagram}
\end{figure}

Finding a general relationship between $a$, $m_c$, and $\Delta \mu$ is challenging because of the averaging procedure we carried out in Section \ref{subsec:astrometry_constraints}. However, the simplifying assumptions that $e \sim i \sim 0$ allow us to reach analytical expressions that still capture the shape of the astrometry posterior. Figure \ref{fig:astro_short_long_diagram} shows a diagram of this scenario. We reproduce Equations \ref{eq:avg_pm} and \ref{eq:dmu_final} below.

\begin{equation}
    \begin{aligned}
        \langle \dot{X} \rangle =  \frac{-1}{t_2 - t_1} \Big [ a_{\star}(\cos E - e) \Big ] ^{E(t_2)} _{E(t_1)} \\ \\
        \langle \dot{Y} \rangle =  \frac{-1}{t_2 - t_1} \Big [ a_{\star}\sqrt{1-e^2} \sin E \Big ]^{E(t_2)} _{E(t_1)}
    \end{aligned}
\label{eq:avg_pm_appendix}
\end{equation}

\begin{gather}
   \Delta \mu = |\vec{\mu}_{G, abs} - \vec{\mu}_{HG, abs}| = |\vec{\mu}_{G, anom} - \vec{\mu}_{HG, anom}|
\label{eq:dmu_final_appendix}
\end{gather}

We begin with the long-period regime, that is, separations corresponding to orbital periods much longer than $T_{HG}$, the 24.75-year interval between the $Hipparcos$ and $Gaia$ characteristic epochs. From Equation \ref{eq:dmu_final_appendix} we see that $\Delta \mu$ depends only on the anomalous proper motions, so we may neglect center-of-mass proper motion in this derivation.

Because the companion orbit is circular and face-on, the proper motion vector has the same magnitude at all epochs, that is, $|\vec{\mu}_H| = |\vec{\mu}_G| = \frac{2 \pi a_{\star}}{d_{\star} P}$. Furthermore, in the long-period regime, the average proper motion approximates the instantaneous value: $|\vec{\mu}_{HG}| \sim |\vec{\mu}_G|$. The proper motion components at the $Gaia$ and $HG$ epochs are thus

\begin{gather}
\begin{aligned}
    \mu_{G_x} = -|\vec{\mu}_G| \sin (2\theta_{HG}) \\
    \mu_{G_y} = |\vec{\mu}_G| \cos (2\theta_{HG})
    \\ \\
    \mu_{HG_x} = -|\vec{\mu}_{HG}| \sin (\theta_{HG}) \\
    \mu_{HG_y} = |\vec{\mu}_{HG}| \cos (\theta_{HG}),
\end{aligned}
\end{gather}

\noindent where $\theta_{HG}$ is the position angle of the host star at the $HG$ epoch (see Figure \ref{fig:astro_short_long_diagram}, left panel), given by

\begin{gather}
\begin{aligned}
    \theta_{HG} = 2 \pi \frac{T_{HG}}{2 P}
    = \frac{T_{HG}}{2} \sqrt{\frac{G(m_{\star}+m_c)}{a^3}}.
\end{aligned}
\end{gather}

From these proper motions, we may compute $\Delta \mu$:

\begin{gather}
\begin{aligned}
    \Delta \mu_{x} &= \mu_{HG} \left [ \sin (\theta_{HG}) - \sin (2\theta_{HG}) \right ] \approx \mu_{HG} \left [ -\theta_{HG} \right ]
    \\
    \Delta \mu_{y} &= \mu_{HG} \left [ \cos (2\theta_{HG}) - \cos (\theta_{HG}) \right ] \approx \mu_{HG} \left [ -\frac{3}{2} \theta_{HG}^2 \right ]
    \\
    \Delta \vec{\mu} &= -\mu_{HG} \left [ \theta_{HG}, \frac{3}{2} \theta_{HG}^2 \right ],
\end{aligned}
\end{gather}

\noindent where we approximated that $\sin \theta_{HG} \sim \theta_{HG}$ and $\cos \theta_{HG} \sim 1-\frac{\theta_{HG}^2}{2}$. We further approximate that the x-component of $\Delta \vec{\mu}$ dominates in the long-period regime, yielding:

\begin{gather}
\begin{aligned}
    |\Delta \vec{\mu}| &\approx \mu_{HG} \cdot \theta_{HG} \\
    &= \frac{2 \pi a_{\star}}{d_{\star} P} \cdot \frac{T_{HG}}{2} \sqrt{\frac{G(m_{\star}+m_c)}{a^3}} \\
    &= \frac{T_{HG}}{2 d_{\star}} a_{\star} \frac{G(m_{\star}+m_c)}{a^3} \\
    &= \frac{G \cdot T_{HG}}{2 d_{\star}} \frac{m_c}{a^2} \\
    &\propto \frac{m_c}{a^2}.
\end{aligned}
\end{gather}

\begin{figure}[!h]
    \centering
    \includegraphics[width=.5\linewidth]{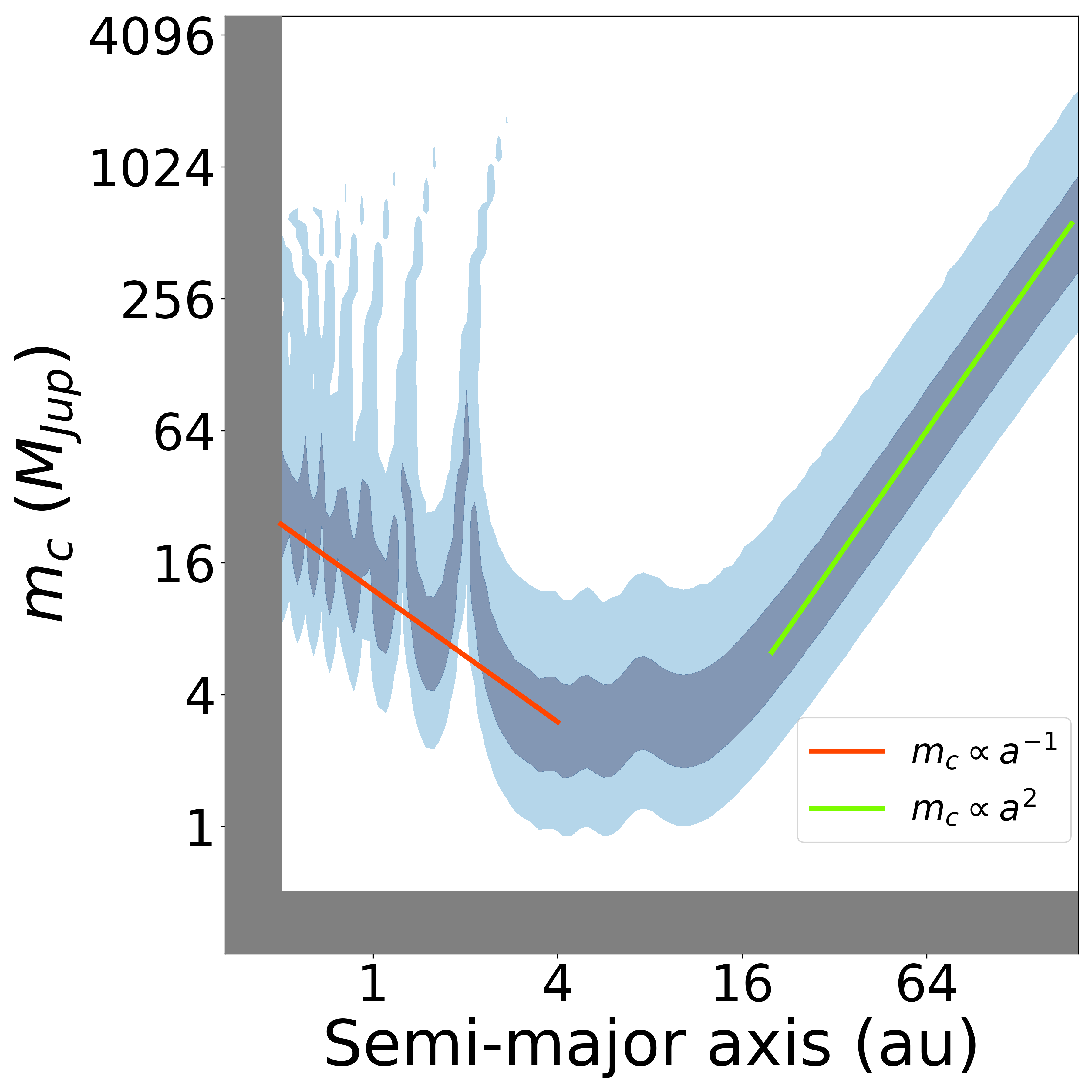}\hfill
    \includegraphics[width=.5\linewidth]{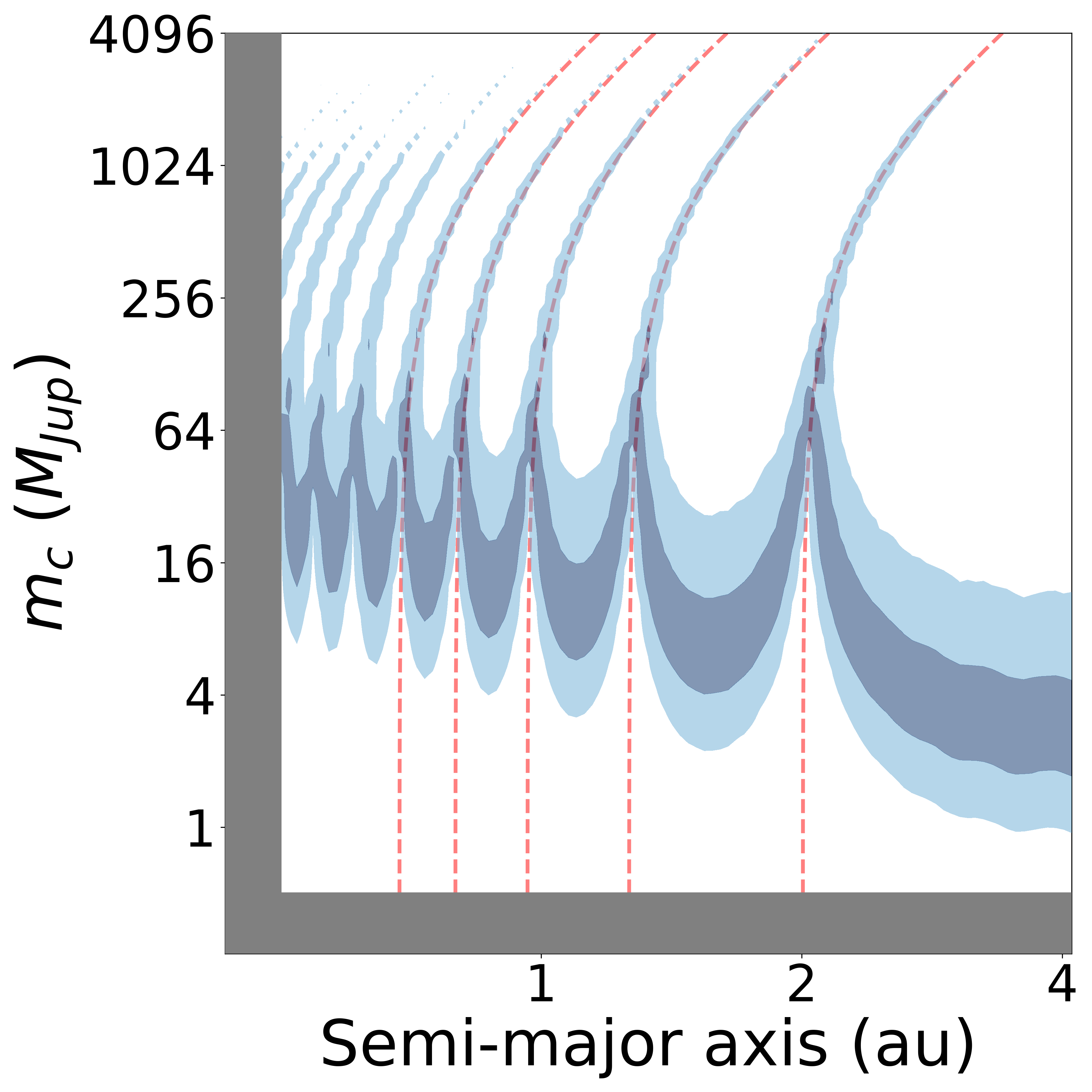}
    \caption{\textbf{Left:} The posterior surface calculated from astrometric data. The red line follows the relationship $m_c \propto a^{-1}$, while the green line follows $m_c \propto a^{2}$. \textbf{Right:} The short-period regime of the surface at left. Iso-period lines are shown in red: the right-most line traces $P=T_{G}\sim2.83$ yr, the next follows $P=\frac{T_{G}}{2}$, then $P=\frac{T_{G}}{3}$, and so on. High-likelihood models near these periods must have large companion masses to counter the small net displacement of the star. The same effect can be seen at harmonics of the $\sim25$-yr \textit{Hipparcos-Gaia} baseline ($\sim 4-16$ AU), though it is less pronounced because $\vec{\mu}_G$ remains nonzero in this regime even when $\vec{\mu}_{HG}$ vanishes.}
    \label{fig:astro_shapes}
\end{figure}

\noindent Thus, $m_c \propto a^2$ in the long-period regime. This behavior is shown for large $a$ values ($a \gtrsim 20$ AU) in the left panel of Figure \ref{fig:astro_shapes}.

The short-period portion of the astrometry posterior surface displays two distinct features: an overall negative slope and a series of spike-like formations extending to high mass. We derive the general linear character of the astrometry posterior in the short-period regime using the same picture we used to understand the long-period behavior. First examining $\vec{\mu}_G$, we have from Equation \ref{eq:avg_pm_appendix}:

\begin{equation}
    \begin{aligned}
    |\vec{\mu}_G| = 
    \frac{1}{d_{\star}}\Bigg | \begin{bmatrix}
                \frac{-a_{\star}}{t_2 - t_1} \cos M \\
                \frac{-a_{\star}}{t_2 - t_1} \sin M
            \end{bmatrix} ^{M(t_2)} _{M(t_1)} \Bigg |,
    \end{aligned}
\label{eq:astro_short_per_dmu}
\end{equation}

\noindent where we have applied our assumptions of a face-on circular orbit and converted velocity to proper motion. Unlike in the long-period regime, $T_G$ is comparable to or multiple times the duration of orbits in the short-period regime. As a result, $M(t_1)$ and $M(t_2)$ may differ by any angle from $0$ to $2\pi$. To continue, we set $M(t_1)=0$ without loss of generality, and average over the possible values of $M(t_2)$:

\begin{equation}
    \begin{aligned}
    |\vec{\mu}_G| d_{\star}
    & \approx
    \Bigg | \begin{bmatrix}
                \frac{-a_{\star}}{t_2 - t_1} \langle \cos M - 1 \rangle \\
                \frac{-a_{\star}}{t_2 - t_1} \langle \sin M - 0 \rangle
            \end{bmatrix} \Bigg |
    &= \frac{a_{\star}}{t_2 - t_1},
    \end{aligned}
\label{eq:astro_short_per_mu_g}
\end{equation}

\noindent where $\langle \quad \rangle$ denotes the average over the interval $\left [ 0 , 2\pi \right ]$. We may apply this same procedure to $\vec{\mu}_{HG}$, but note that in this case, the denominator of the right-hand side of Equation \ref{eq:astro_short_per_mu_g} is $\sim25$ years, as opposed to $\lesssim 3$ years for $\vec{\mu}_G$. Thus, $\vec{\mu}_G$ dominates in the short-period regime, and we neglect $\vec{\mu}_{HG}$. See the right panel of Figure \ref{fig:astro_short_long_diagram} for a pictorial representation of this scenario. Finally, we use the relation between $a$ and $a_{\star}$ to find that $m_c \propto a^{-1}$ for short periods:

\begin{equation}
    |\vec{\mu}_G| d_{\star} \approx \frac{a_{\star}}{t_2 - t_1} = \frac{a}{t_2 - t_1} \frac{m_c}{m_{\star} + m_c} \propto a \cdot m_c.
\label{eq:astro_short_per_relation}
\end{equation}

\noindent We show this relation in Figure \ref{fig:astro_shapes}.

The spikes represent harmonics of $T_{G}$, the 34-month $Gaia$ EDR3 mission baseline. For periods shorter than the $\sim25$-year \textit{Hipparcos-Gaia} baseline, averaging the orbital position over that interval produces a cancellation effect, reducing the magnitude of $\vec{\mu}_{HG}$ so that $\vec{\mu}_{G}$ dominates $\Delta \mu$ in the short-period regime. For orbital periods which divide $T_{G}$, i.e. $T_{G}$ itself, $\frac{T_{G}}{2}$, etc., $\vec{\mu}_{G} \to 0$ as well because the host star returns to its original position over the course of the $Gaia$ mission, resulting in an average measured velocity of zero. The spikes in the right panel of Figure \ref{fig:astro_shapes} demonstrate that model orbits must have higher companion mass near the harmonics to produce a constant value of $\Delta \mu$. We observe a similar effect at harmonics of $T_{HG}$ (left panel of Figure \ref{fig:astro_shapes}), though these spikes are less pronounced because both $\vec{\mu}_{HG}$ and $\vec{\mu}_{G}$ contribute to $\Delta \mu$ in that regime.

\section{Appendix: Rotation Matrix}
\label{appendix:rot_matrix}
\textbf{R} is the rotation matrix \textbf{P$_3$P$_2$P$_1$}, described in Equations 2.119-2.121 of \cite{murray_dermott_2010}. It is not given explicitly in that text.

\begin{gather}
     \textbf{R} = 
     \begin{pmatrix}
      \cos \Omega \cos \omega - \cos i \sin \Omega \sin \omega & -\cos \Omega \sin \omega -  \cos i \sin \Omega \cos \omega &  \sin i \sin \Omega \\
      \sin \Omega \cos \omega + \cos i \cos \Omega \sin \omega & \cos i \cos \Omega \cos \omega - \sin \Omega \sin \omega & -\sin i \cos \Omega \\
      \sin i \sin \omega & \sin i \cos \omega & \cos i 
    \end{pmatrix}
\end{gather}

We exploit a symmetry in Sections \ref{subsec:astrometry_constraints} and \ref{subsec:imaging_constraints} which warrants explanation here. In those sections, we apply \textbf{R} to the \textit{stellar} position and velocity vectors, despite the fact that $i$, $\omega$, and $\Omega$ correspond to the \textit{companion} orbit. $\Omega$ and $i$ are the same for the star and the companion, but the stellar argument of periapsis is offset from that of the companion by $\pi$ radians: $\omega_{\star} = \omega_c + \pi$. This means that rotating the stellar position and velocity vectors using $\omega_c$ is not correct in general. We demonstrate below that for our calculations, this asymmetry vanishes.

First, we relabel the matrix \textbf{R} above as \textbf{R$_c$} to denote its use of the sampled $\omega_c$. We also define the matrix \textbf{R$_{\star}$}, which results from the substitution $\omega_c \rightarrow \omega_{\star}$:

\begin{equation}
    \begin{aligned}
             \textbf{R$_c$} = 
                \begin{pmatrix}
                R_{0,0} & R_{0,1} & R_{0,2} \\
                R_{1,0} & R_{1,1} & R_{1,2} \\
                R_{2,0} & R_{2,1} & R_{2,2} 
                \end{pmatrix}, \quad
            \textbf{R$_{\star}$} = 
                \begin{pmatrix}
                -R_{0,0} & -R_{0,1} & R_{0,2} \\
                -R_{1,0} & -R_{1,1} & R_{1,2} \\
                -R_{2,0} & -R_{2,1} & R_{2,2} 
                \end{pmatrix}.
    \end{aligned}
\end{equation}

\noindent where we have abbreviated the matrix elements for clarity.

We summarize the rotation steps of Section \ref{subsec:astrometry_constraints} as

\begin{equation}
    \begin{aligned}
        \vec{r}_{obs} = \boldsymbol{R}_c \, \vec{r}_{orb} =  
        \boldsymbol{R}_c
            \begin{bmatrix}
                \langle X \rangle \\
                \langle Y \rangle \\
                0
            \end{bmatrix} = 
            \begin{bmatrix}
                R_{0,0}\langle X \rangle + R_{1,0}\langle Y \rangle \\
                R_{1,0}\langle X \rangle + R_{1,1}\langle Y \rangle \\
                0
            \end{bmatrix} = -
            \begin{bmatrix}
                -R_{0,0}\langle X \rangle + (-R_{1,0}\langle Y \rangle) \\
                -R_{1,0}\langle X \rangle + (-R_{1,1}\langle Y \rangle) \\
                0
            \end{bmatrix} = -
        \boldsymbol{R}_{\star}
            \begin{bmatrix}
                \langle X \rangle \\
                \langle Y \rangle \\
                0
            \end{bmatrix}.
            \\ \\
    \end{aligned}
\end{equation}

The situation is analogous for $\vec{v}_{obs}$, as well as for $\vec{\rho}$ in Section \ref{subsec:imaging_constraints}. Thus, the use of \textbf{R$_{c}$} instead of \textbf{R$_{\star}$} introduces a single negative sign to each rotated vector. Because all vectors are equally affected, and in all cases we ultimately work with the vector moduli, this negative sign has no effect on our likelihood calculation.

\end{document}